\documentclass[journal]{IEEEtran}
\usepackage{amsmath,amsfonts,amssymb}
\usepackage{algorithmic}
\usepackage{algorithm}
\usepackage{array}
\usepackage[caption=false,font=scriptsize,labelfont=rm,textfont=rm]{subfig}
\usepackage{textcomp}
\usepackage{url}
\usepackage{verbatim}
\usepackage{graphicx}
\usepackage{cite}
\usepackage{xcolor}
\usepackage{titlesec}
\usepackage{hyperref}
\hypersetup{
    bookmarks=true,         
    colorlinks=true,       
    linkcolor=blue,          
    citecolor=blue,
    filecolor=cyan,         
    urlcolor=magenta        
}

\usepackage{bm} 
\usepackage{amsmath,amssymb}
\usepackage{booktabs} 
\usepackage{multirow}
\usepackage{framed}
\usepackage{tabularx}
\usepackage{array}
\usepackage{enumitem}
\usepackage[dvipsnames]{xcolor} 
\usepackage[framemethod=TikZ]{mdframed}

\newcommand{\ut}[1]{\bm{u}_{#1}}

\newcommand{\uttp}[1]{\tilde{\bm{u}}^\prime_{#1}}
\newcommand{\utp}[1]{\bm{u}^\prime_{#1}}

\newcommand{\uts}[1]{\dot{\bm{u}}_{#1}}
\newcommand{\vt}[1]{\bm{v}_{#1}}
\newcommand{\vtt}[1]{\tilde{\bm{v}}_{#1}}
\newcommand{\xt}[1]{\bm{x}_{#1}}

\newcommand{\lambdar}[0]{\lambda_{\text{r}}}

\newcommand{\U}[0]{\textrm{U}}
\renewcommand{\L}[0]{\textrm{L}}

\newcommand{\norm}[1]{\lVert#1\rVert}
\newcommand{\ceil}[1]{\lceil#1\rceil}

\newcommand{\ebp}[0]{\epsilon_{\textrm{BP}}}

\usepackage{calc}
\makeatletter
\def\textSq#1{%
	\begingroup
	\setlength{\fboxsep}{0.3ex}
	\setbox1=\hbox{#1}
	\setlength{\@tempdima}{\maxof{\wd1}{\ht1+\dp1}}
	\setlength{\@tempdimb}{(\@tempdima-\ht1+\dp1)/2}
	\raise-\@tempdimb\hbox{\fbox{\vbox to \@tempdima{%
				\vfil\hbox to \@tempdima{\hfil\copy1\hfil}\vfil}}}%
	\endgroup%
}
\makeatother

\usepackage{color} 

\begin{document}

\title{Recent Advances in Spatially Coupled Codes: Overview and Outlook}

\author{Min Qiu,~\IEEEmembership{Senior Member,~IEEE}, Xiaowei Wu,~\IEEEmembership{Member,~IEEE}, Peng Kang,~\IEEEmembership{Senior Member,~IEEE}, Lei Yang,~\IEEEmembership{Member,~IEEE}, Jinhong Yuan,~\IEEEmembership{Fellow,~IEEE}
\thanks{Corresponding authors: Xiaowei Wu and Peng Kang.
}
}

\markboth{IEEE BITS the Information Theory Magazine}%
{Shell \MakeLowercase{\textit{et al.}}: A Sample Article Using IEEEtran.cls for IEEE Journals}


\maketitle

\begin{abstract}
The concept of spatial coupling is among the most significant breakthroughs in coding theory over the past decade. The excellent waterfall and error floor performance of spatially coupled codes has positioned them as promising coding candidates for future communication and data storage systems. This article presents an overview of recent advances in spatially coupled codes. In particular, we first review several representative examples of recently proposed spatially coupled codes and highlight their unique features that make them appealing for different applications. Next, we discuss the useful properties of spatially coupled codes and how to design good spatially coupled codes. The article concludes with some future research directions and open problems.
\end{abstract}


\section{Introduction}
Channel coding is essential in communication systems to ensure reliable and efficient communications. Since Shannon published his landmark paper in 1948, significant efforts have been made to construct codes that can approach capacity with reasonable complexity. Most notably, the invention of turbo codes and polar codes, as well as the rediscovery of low-density parity-check (LDPC) codes, have made this goal a reality. Due to their capacity-approaching performance with moderate decoding complexity, many applications, such as cellular networks, Wi-Fi, and satellite communications, have adopted these codes as industry standards.

These successes have triggered more interest in the research of more powerful channel codes. Aiming to further improve the performance of conventional linear block codes, researchers have turned their attention to convolutional-like codes. The earliest work in this direction is \cite{782171}, which introduced time-varying LDPC convolutional codes with large memory. These constructions are now understood as instances of spatially coupled LDPC (SC-LDPC) codes.

SC-LDPC codes possess several unique and powerful features that set them apart from conventional LDPC block codes. Despite having almost regular node degrees, except for a slight irregularity at the ends of the chain due to termination, regular SC-LDPC codes can achieve significantly better thresholds than their uncoupled block counterparts \cite{5571910}. Interestingly, it was numerically observed in \cite{5571910} that the belief propagation (BP) decoding threshold (BP threshold) of regular SC-LDPC codes coincides with the \emph{maximum a posteriori} probability (MAP) decoding threshold (MAP threshold) of the underlying uncoupled codes. This phenomenon is referred to as threshold saturation (see Sec. \ref{sec:thres_sa} for a detailed discussion) and the first analytical proof for transmission over the binary erasure channel (BEC) was given in \cite{5695130}. The implication of this result is that spatial coupling enables the underlying codes to achieve excellent performance with low-complexity sub-optimal decoding algorithms.

Furthermore, \cite{5571910,5695130} show that increasing the variable and check node degrees of regular SC-LDPC codes gradually pushes the BP thresholds to the capacity. This eliminates the need for meticulous optimization of their degree distributions, as required for conventional LDPC block codes. In addition, the minimum distance of regular SC-LDPC codes has been shown to grow linearly with blocklength, promising good error floor performance. Last but not least, they can be decoded by using sliding window decoding with manageable decoding latency and complexity. As a result of these features, spatial coupling provides an effective framework to construct powerful channel codes.

The research on spatial coupling remains active. Motivated by the success of SC-LDPC codes, researchers have applied spatial coupling to various classes of codes tailored to different applications. For example, in \cite{Smith12}, a class of spatially coupled product codes with Bose–Chaudhuri–Hocquenghem (BCH) component codes, dubbed staircase codes, were introduced for high-speed optical communications and adopted in relevant standards, e.g., OIF 400ZR \cite{400ZR}. Staircase codes demonstrate excellent waterfall (within 0.56 dB to the binary symmetric channel (BSC) capacity) and error floor performance (bit error rate below $10^{-15}$) under iterative hard-decision decoding at high rates.

It is also important to realize that future generation communication systems are expected to support diverse communication scenarios and channel conditions. Explicitly designing and optimizing a particular code for each scenario and channel condition is deemed impractical. In this regard, it would be desirable to design \emph{universal} codes that are simultaneously good for different channels, meeting the diverse demands of future communication systems. For regular SC-LDPC codes, it was proven that threshold saturation occurs for general binary memoryless symmetric (BMS) channels \cite{6912949}. Consequently, a single regular SC-LDPC code ensemble is capable of achieving the capacity of the whole class of BMS channels \emph{universally} under BP decoding \cite{6589171}. This suggests that spatial coupling may hold the key to constructing universal codes.

Yet, many theoretical results on spatially coupled codes were established based on some ideal assumptions, such as randomized coding structures, infinite length, coupling memory, decoding iterations, etc. To this end, recent research on spatially coupled codes has focused on addressing their practical limitations and achieving a better tradeoff between performance, complexity, and latency. In this article, we will provide an overview of the recent advances in the theory and practice of spatially coupled codes. We conclude by highlighting several open challenges and future research directions. We hope that this article will provide guidance towards future coding designs and trigger new research in spatially coupled coding structures and spatial coupling beyond channel coding.

The remainder of this article is organized as follows. Sec. \ref{sec:2} reviews the main classes of spatially coupled codes. Sec. \ref{sec:3} discusses their key properties. Sec. \ref{sec:4} focuses on the design aspects of spatially coupled codes. Sec. \ref{sec:5} provides numerical results of representative spatially coupled codes. Finally, Sec. \ref{sec:6} outlines open problems and future research directions.

\section{Classes of Spatially Coupled Codes}\label{sec:2}
In this section, we review several representative examples of recently proposed spatially coupled codes, highlighting their distinctive features and potential impact on supporting different applications. We begin by providing a general description of spatially coupled codes.

\begin{figure}[t]
    \centering
    \includegraphics[width=7cm]{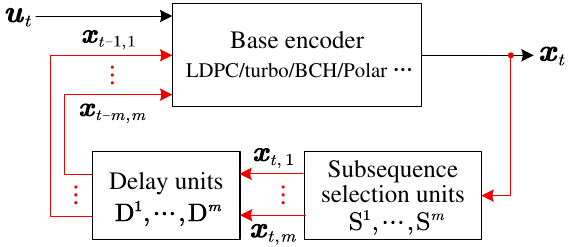}
 \caption{General structure of spatially coupled codes. For a coupling memory $m$ and $i\in\{1,\ldots,m\}$, the selection unit $\text{S}^{i}$ extracts a (sub)sequence $\xt{t,i}$ from $\xt{t}$. $\text{D}^{i}$ denotes a delay unit that holds the input for $i$ time instant(s).}\label{fig:SC_gen_encoder}
\end{figure}

\begin{framed} %
             \noindent\textbf{General Template of a Spatially Coupled Code}

            A spatially coupled code can be defined by the following parameters and components:
            \begin{itemize}
                \item \emph{Base code}: An $(N,K)$ code that maps a length-$K$ information sequence to a length-$N$ codeword.
                \item \emph{Coupling length} $L$: The number of coupled base code blocks.
                \item \emph{Coupling memory} $m$ (or coupling width): The maximum number of preceding time instants that contribute to the generation of the current codeword.
    \end{itemize}
\end{framed}
	
The encoding process of a spatially coupled code with coupling length $L$ and coupling memory $m$ can be described as follows:
\begin{enumerate}
        \item Initialize $\xt{t}=\boldsymbol{0}$ for $t=-m+1,\dots,0$.
        \item At time $t=1,\ldots,L$, as depicted in Fig. \ref{fig:SC_gen_encoder}, the encoding proceeds as follows:
        \begin{enumerate}
            \item The base encoder takes a length-$K$ input block, consisting of information sequence $\ut{t}$ and $m$ sequences $\xt{t-1,1},\ldots,\xt{t-m,m}$ originating from the previously generated codewords. It then produces a length-$N$ component codeword $\xt{t}$.
            \item Extract $m$ (sub)sequences $\xt{t,1},\ldots,\xt{t,m}$ from $\xt{t}$.
            \item For $i=1,\ldots,m$, delay $\xt{t,i}$ by $i$ time instants. This delayed sequence is then used as a coupling input to the base encoder at time $t+i$.
        \end{enumerate}
        \item Set $\ut{t}=\boldsymbol{0}$ for $t = L-m+1,\ldots,L$ to terminate the coupling.
\end{enumerate}
\noindent Here, $\xt{t-1,1},\ldots, \xt{t-m,m}$ are referred to as the \emph{coupling inputs} to the base encoder at time $t$. Within this architecture, the base encoder could take various forms, such as a single block encoder, multiple concatenated component encoders, or a set of parallel component encoders. Unless otherwise specified, all spatially coupled codes considered in this article are \emph{terminated}.

Spatially coupled codes can be decoded by performing full decoding over the entire coupled chain.
In practice, a sliding window decoder is used to reduce the decoding latency \cite{6086762}. Consider a decoding window size of $W$. The decoding operates as follows:%
\begin{enumerate}
        \item Initialize the window position at $t=1$.
        \item Conduct iterative decoding on the $W$ codewords $\xt{t}, \dots, \xt{t+W-1}$ that span the current window. 
        \item Once the iterations are complete or a predefined stopping criterion is met, output the estimated information sequence $\hat{\bm{u}}_{t}$.
        \item Slide the decoding window forward by one block, i.e., assign $t \leftarrow t+1$.
        \item Repeat steps 2 - 4 until all information sequences are decoded.
\end{enumerate}%
A decoding wave is triggered by the known information introduced through termination of the coupling chain, often referred to as ``seeding'' at the boundaries. Through this wave, reliable information is propagated across consecutive decoding windows.

\subsection{Spatially Coupled LDPC Codes}\label{sec:scldpc}
SC-LDPC codes are the canonical example that shows threshold saturation and admit graph-based constructions. An SC-LDPC code with coupling memory $m$ and coupling length $L$ can be described based on Fig. \ref{fig:SC_gen_encoder}, where the base encoder is the LDPC block encoder. At time $t$, the LDPC block encoder takes the coupling inputs $\xt{t-1}, \ldots, \xt{t-m}$ together with $\boldsymbol{u}_t$ to generate component codeword $\xt{t}$. Moreover, the sequences $\xt{t}, \xt{t-1}, \ldots, \xt{t-m}$ are jointly involved in the same check equation, e.g., \cite[Eq. (4)]{5571910}.

For initialization, $\xt{t}$ is set to $\boldsymbol{0}$ for $t<1$.
To terminate the transmission after coupling $L$ codewords, the information sequence $\ut{t}$ is set to $\boldsymbol{0}$ for $t > L$, yielding $\xt{t} = \boldsymbol{0}$ for $t>L$.
A sliding window decoder operates as described in Sec. \ref{sec:2}, wherein iterative decoding, such as BP decoding, is performed within each window.

In general, SC-LDPC code ensembles can be categorized as follows:
\begin{itemize}
    \item \emph{Randomized ensembles:} Edge connections are assigned randomly. These ensembles are primarily used for analysis, e.g., \cite{5695130,6589171,6912949}.
    \item \emph{Structured ensembles:} Edge connections are subject to predefined structural constraints. These ensembles are designed for efficient implementation, e.g., \cite{Mitchell2015sc}.
\end{itemize}

\begin{figure}[t!]
	\centering
	\includegraphics[width=\linewidth]{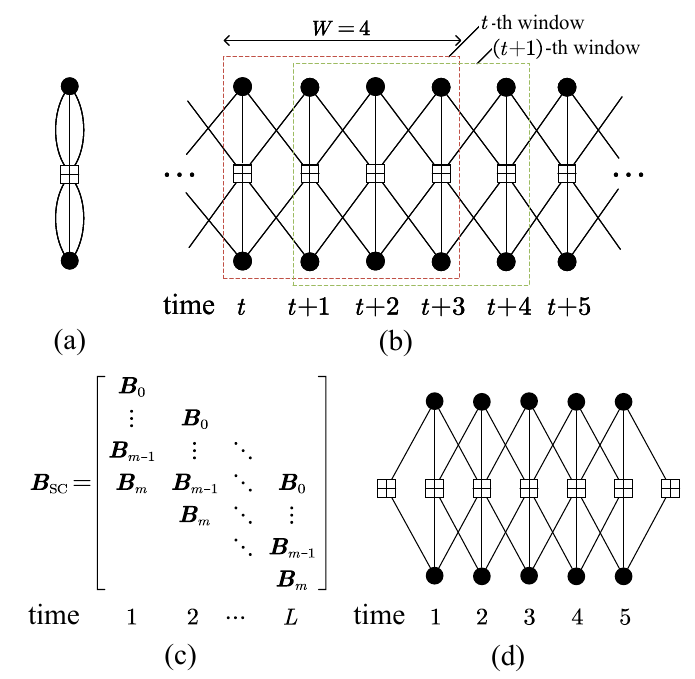}
	\caption{(a) An uncoupled protograph for $\boldsymbol{B}_{\text{BC}} = [3\,\,3]$, representing a $(3,6)$-regular LDPC block code. (b) The spatially coupled chain of protographs across time instants, with an illustration of a sliding window decoder of size $W=4$. (c) The base matrix of a time-invariant SC-LDPC protograph with coupling length $L$. (d) The SC-LDPC protograph with $L=5$ and $m=2$. $\boxplus$ denotes a CN and $\bullet$ denotes a VN in the protograph.}
	\label{fig:termination}
\end{figure}

A representative example of structured SC-LDPC code ensembles is the time-invariant protograph-based SC-LDPC code ensemble \cite{Mitchell2015sc}. Let $d_{\text{v}}$ and $d_{\text{c}}$ denote the variable node (VN) and check node (CN) degrees, respectively. The construction of a terminated protograph-based  $(d_{\text{v}},d_{\text{c}})$-regular SC-LDPC code with coupling length $L$ and coupling memory $m$ is described as follows:
\begin{enumerate}
    \item \textit{Construct the base code}: Start from a protograph-based $(d_{\text{v}},d_{\text{c}})$-regular LDPC block code described by a $b_{\text{c}} \times b_{\text{v}}$ base matrix $\boldsymbol{B}_{\text{BC}}$, which defines an uncoupled protograph with $b_{\text{c}}$ CNs and $b_{\text{v}}$ VNs (see Fig. \ref{fig:termination}(a) for an example).
    \item \textit{Replicate in time}: Replicate the uncoupled protograph derived from $\boldsymbol{B}_{\text{BC}}$ across time instants $t=1,2,\ldots$.
    \item \textit{Apply edge spreading}: Apply \textit{edge spreading} to the uncoupled protographs (see Fig. \ref{fig:termination}(b)). This operation decomposes $\boldsymbol{B}_{\text{BC}}$ into $m+1$ component base matrices $\boldsymbol{B}_{\text{BC}} = \sum^{m}_{i=0}\boldsymbol{B}_i$, where $\boldsymbol{B}_i$ specifies the edge connections between $b_{\text{v}}$ VNs at the $t$-th time instant and $b_{\text{c}}$ CNs at the $(t+i)$-th time instant, thereby encoding connections across time instants separated by $i$.
    \item \textit{Terminate the chain}: Truncate the coupled chain after $L$ time instants to obtain a terminated $(d_{\text{v}},d_{\text{c}})$-regular SC-LDPC protograph over time instants $t = 1,2,\ldots,L$, whose base matrix $\boldsymbol{B}_{\text{SC}}$ has a banded structure (see Fig. \ref{fig:termination}(c)), with Each column corresponding to a time instant.
    \item \textit{Lift to obtain the parity-check matrix}: Replace each nonzero entry in $\boldsymbol{B}_{\text{SC}}$ with an $M\times M$ circulant permutation matrix, where $M$ denotes the \textit{lifting factor}.
\end{enumerate}
The resulting protograph-based $(d_{\text{v}},d_{\text{c}})$-regular SC-LDPC code has a design rate of $R = 1-\frac{(L+m)b_{\text{c}}}{Lb_{\text{v}}}$.

The protograph-based spatially coupled ensembles differ from the randomized spatially coupled ensembles in \cite{5695130,6589171,6912949} in the following aspects:
\begin{itemize}
\item \emph{Randomized ensembles:} Assign a \textit{nonzero} probability to \textit{all possible} edge spreading types at each position in the chain, resulting in a \textit{nonzero} fraction of VNs at every spatial position exhibiting each such edge spreading type.
\item \emph{Protograph-based ensembles:} Restrict the construction to a specific (and typically small) number of edge spreading types at each position.
\end{itemize}
As noted in \cite[Sec. II-E3]{Mitchell2015sc}, the randomized SC-LDPC code ensemble \cite{5695130,6589171} does not include any specific protograph-based code ensemble.

\medskip
\noindent\textbf{Example:}
Fig. \ref{fig:termination}(d) shows a spatially coupled protograph constructed from a (3,6)-regular LDPC block code with $L=5$ and $m=2$.
The edge spreading decomposes the underlying $\boldsymbol{B}_{\text{BC}}$ into $\boldsymbol{B}_i = [1, 1]$ for $i=0,1,2$, leading to a design rate of 0.4 for this SC-LDPC code ensemble.
The rate loss comes from termination and becomes negligible as $L$ increases.

Finally, note that replacing some or all single parity-check nodes in the graph by generalized constraint nodes yields spatially coupled generalized LDPC codes \cite{9398939}.

	\subsection{Spatially Coupled Turbo-Like Codes} \label{sec:sctc}
Spatial coupling has been applied to codes on graphs whose constraint nodes are convolutional codes (CCs), resulting in spatially coupled turbo-like codes (SC-TCs) \cite{Moloudi-scTurbo}. SC-TCs can be described by the structure shown in Fig. \ref{fig:SC_gen_encoder}, where the base codes are concatenated CCs. The coupling pattern of an SC-TC depends on the specific code construction and is described next. The termination of SC-TCs follows the procedure as that used for the codes shown in Fig. \ref{fig:SC_gen_encoder}. SC-TCs can be decoded using a sliding window decoder, where the decoding of each component codeword $\boldsymbol{x}_t$ involves inner turbo iterations, with the constituent convolutional codes decoded using the Bahl–Cocke–Jelinek–Raviv (BCJR) algorithm \cite{1055186}.

In what follows, we review three representative examples of SC-TCs constructed respectively using the following base codes:
    \begin{itemize}
        \item \emph{Parallel concatenated codes (PCCs):} The component encoders operate in parallel on the same information sequence;
        \item \emph{Serially concatenated codes (SCCs):} The component encoders are connected in series through an interleaver;
        \item \emph{Braided convolutional codes (BCCs):} The component encoders are braided by feeding back parity bits as inputs for subsequent encoding time instants.
    \end{itemize}

        \begin{figure}[t!]
	\centering
\includegraphics[width=0.9\linewidth]{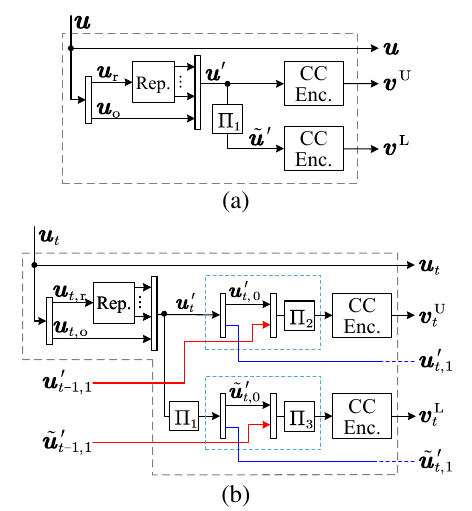}
\caption{Block diagram of (a) a PR-PCC encoder and (b) a GSC-PCC encoder with coupling memory $m=1$. The narrow rectangular  \raisebox{-1.5pt}{\scalebox{0.4}[1.3]{$\square$}} with arrows connected refers to a multiplexer or demultiplexer, and $\boxed{\Pi}$ denotes an interleaver. The light-blue boxes highlight the coupling operations.}
\label{fig:gscpcc}
\end{figure}
	
\vspace{0.5em}	
\subsubsection{Spatially Coupled PCCs}	
PCCs, or turbo codes, are the most widely studied class of turbo-like codes and have been adopted in various communication standards. Spatially coupled PCCs (SC-PCCs) \cite{Moloudi-scTurbo} offer significant performance gain over PCCs while maintaining potential compatibility with existing standard PCCs. However, their BP threshold remains strictly bounded away from capacity \cite[Table II]{Moloudi-scTurbo}, especially after puncturing. To close this gap, generalized spatially coupled PCCs (GSC-PCCs), which incorporate partial repetition of the information bits, can achieve capacity on the BEC \cite{min_gscpcc}.

GSC-PCCs can be described by the block diagram in Fig. \ref{fig:SC_gen_encoder}, where the base code is a PCC with partial information repetition (PR-PCC) shown in Fig. \ref{fig:gscpcc}(a). In the PR-PCC, a subsequence $\ut{\text{r}}$ of the information sequence $\ut{}$ is repeated, resulting in a partially repeated sequence $\ut{}^\prime$. This repetition is characterized by:
    \begin{itemize}
        \item \emph{Repetition factor} $q$: The number of times the subsequence $\ut{\text{r}}$ is repeated.
        \item \emph{Repetition ratio} $\lambda_{\text{r}}$: The ratio of the length of $\ut{\text{r}}$ to that of $\ut{}^\prime$, i.e. $\lambda_{\text{r}} = \frac{\norm{\ut{\text{r}}}} {\norm{\ut{}^\prime}} \in(0,\frac{1}{q}]$, where $\| \cdot \|$ returns the  vector length.
\end{itemize}
\noindent Next, $\ut{}^\prime$ and its permuted copy $\tilde{\boldsymbol{u}}'$ are encoded by the upper and lower CC encoders, respectively, giving the codeword $\xt{}=[\ut{},\vt{}^\U,\vt{}^\L]$, where $\vt{}^\U$ and $\vt{}^\L$ denote the upper and lower parity sequences, respectively. 	

Intuitively speaking, similar to repeat-accumulate codes \cite{divsalar1998coding}, the repetition provides stronger protection for the repeated bits. GSC-PCCs leverage both spatial coupling and repetition to further amplify the performance gains by propagating the highly protected bits across the coupled structure and enhancing the reliability of neighboring components.
Most significantly, GSC-PCCs are proven to have threshold saturation and, to the best of our knowledge, constitute the first class of SC-TCs proven to be capacity-achieving on the BEC \cite[Sec. IV]{min_gscpcc}.

The GSC-PCC has symmetric encoding operations at the upper and lower CC encoders. Hence, we only describe the coupling process at the upper CC encoders, as the lower encoders operate analogously. At time $t=1,\dots,L$, the upper CC encoding proceeds as follows:
\begin{enumerate}
    \item Partition the repeated sequence $\utp{t}$ into $\utp{t,0},\dots,\utp{t,m}$.
    \item Collect the sequences $\utp{t-m,m},\dots,\utp{t-1,1}$ from previous $m$ time instants as the coupling inputs.
    \item Interleave $\utp{t,0}$ along with the coupling inputs, and encode the resulting sequence using the upper CC encoder to generate the upper parity $\vt{t}^\U$.
\end{enumerate}
\noindent Likewise, the same operations are applied to the permuted sequence $\uttp{t}$ at the lower CC encoder, producing the lower parity $\vt{t}^\L$. As an example, Fig. \ref{fig:gscpcc}(b) depicts a GSC-PCC encoder with coupling memory $m=1$. Notably, by setting $q=1$, this process gives the conventional SC-PCCs proposed in \cite{Moloudi-scTurbo}.

The code rate of a GSC-PCC$(q,\lambda_{\text{r}})$ is given by $R=\frac{(1-\lambda_{\text{r}} (q-1))(L-m)}{((1/R_0)-\lambda_{\text{r}} (q-1)) L}\overset{\textrm{\scalebox{0.8}{$L\!\rightarrow\!\infty$}}}{=}\frac{1-\lambda_{\text{r}} (q-1)}{(1/R_0)-\lambda_{\text{r}} (q-1)}$, where $R_0$ is the code rate of the mother PCC. The parameters $q$ and $\lambda_{\text{r}}$ follow the same definitions as in PR-PCCs and jointly affect the error performance of GSC-PCCs.

\vspace{0.5em}
\subsubsection{Spatially Coupled SCCs}
SCCs benefit more significantly from spatial coupling than PCCs. Although SCCs exhibit a superior error floor, they suffer from considerably worse decoding thresholds than PCCs. To overcome this limitation, spatially coupled SCCs (SC-SCCs) were introduced in \cite{Moloudi-scTurbo}. These codes achieve substantially better decoding thresholds than the SC-PCCs presented in the same paper and are proven to exhibit threshold saturation \cite[Sec. VIII]{Moloudi-scTurbo}.

The encoder of an SCC consists of an outer encoder and an inner encoder. The outer encoder takes $\ut{}$ to produce the outer codeword $\vtt{}^{\text{O}} = [\ut{}, \vt{}^{\text{O}}]$, where $\vt{}^{\text{O}}$ denotes the outer parity sequence. The outer codeword is used as the input of the inner encoder, resulting in the SCC codeword $\xt{}=[\ut{},\vt{}^{\text{O}},\vt{}^{\text{I}}]$, where $\vt{}^{\text{I}}$ denotes the inner parity sequence.

\begin{figure}[t]
    \centering
    \includegraphics[width=7.2cm]{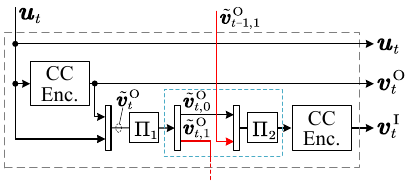}
    \caption{Block diagram of an SC-SCC encoder with rate-$1/2$ convolutional component codes and coupling memory $m=1$.}
    \label{fig:scscc_enc}
\end{figure}

With SCCs as the base code, the encoding of SC-SCCs at time $t \in \{1,\ldots,L\}$ is performed as follows:
\begin{enumerate}
        \item Interleave the outer codeword $\vtt{t}^{\text{O}}$ and partition it into $\vtt{t,0}^{\text{O}},\dots,\vtt{t,m}^{\text{O}}$.
        \item Collect $\tilde{\boldsymbol{v}}_{t-m,m}^{\text{O}}, \dots ,\tilde{\boldsymbol{v}}_{t-1,1}^{\text{O}}$ as the coupling inputs.
        \item Interleave $\tilde{\boldsymbol{v}}_{t,0}^{\text{O}}$ along with the coupling inputs, and encode the resulting sequence using the inner CC encoder to produce the inner parity $\vt{t}^{\text{I}}$.
\end{enumerate}
The code rate of an SC-SCC is $R=\frac{(L-m)R_0}{L}\overset{\textrm{\scalebox{0.8}{$L\!\rightarrow\!\infty$}}}{=}R_0$, where $R_0$ is the rate of the mother SCC. An example of SC-SCC with $m=1$ is depicted in Fig. \ref{fig:scscc_enc}.

Notably, although the outer parity sequences in SCCs and SC-SCCs are usually fully punctured in the literature \cite[Table II]{Moloudi-scTurbo}, our density evolution (DE) results in Table \ref{tb:sctc_de} confirm that transmitting the outer parity sequences in SC-SCCs can improve the BP threshold when $m$ is small compared to the case where they are fully punctured.

\vspace{0.5em}
\subsubsection{Spatially Coupled BCCs}

\begin{figure}[t]
    \centering
    \includegraphics[width=7cm]{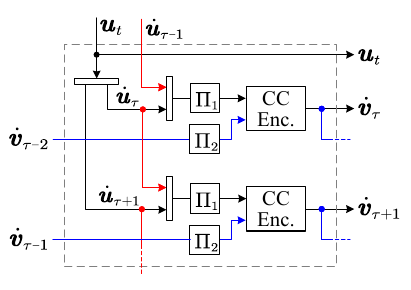}
    \caption{Block diagram of an HSC-BCC encoder with rate-$2/3$ convolutional component codes and coupling memory $m=1$ $(\sigma=2)$.} \label{fig:hscpcc_enc}
\end{figure}

BCCs, introduced in \cite{5361461}, possess an inherent spatially coupled structure, in which the parity sequences generated at time $t$ are fed back as inputs to the encoders at time $t+1$. These codes can be viewed as the convolutional counterparts of braided block codes \cite{4957627} that served as an important precursor to the staircase codes discussed in Sec. \ref{sec:stair}. BCCs and their variants exhibit excellent decoding thresholds and error floor performance.

The work in \cite{Moloudi-scTurbo} generalized this concept to spatially coupled braided convolutional codes (SC-BCCs) by extending the coupling memory to $m > 1$ and by incorporating the coupling methods of SC-PCCs and BCCs. More recently, half spatially coupled BCCs (HSC-BCCs) \cite{xw_hscbcc} were proposed to further improve the decoding threshold while maintaining a small coupling memory.

Before describing the coupling operation of HSC-BCCs, we first introduce the following notation:
\begin{itemize}
\item \emph{Half-time instant:} $\tau \in \{1,...,2L\}$, where $\tau=2t-1$ and $\tau+1=2t$ both correspond to time instant $t$, with the former associated with the upper encoder and the latter with the lower encoder.
\item \emph{Coupling memory in the half-time instant sense:} $\sigma \geq 2$ and $m=\ceil{\sigma /2}$.
\end{itemize}

The coupling in HSC-BCCs is performed at the half-time-instant level. Specifically, for each time instant $t = 1, \ldots, L$, the information sequence $\ut{t}$ is partitioned into two equal-length subsequences, $\uts{\tau}$ and $\uts{\tau+1}$. Let $\dot{\boldsymbol{v}}_{\tau}$ denote the parity sequence at half-time instant $\tau$. The encoding of HSC-BCCs at $\tau$ is performed as follows:
\begin{enumerate}
        \item Collect $\uts{\tau-\sigma+1}$ and $\dot{\boldsymbol{v}}_{\tau-\sigma}$ from the previous half-time instants as the coupling inputs.
                 \item Interleave $[\uts{\tau-\sigma+1},\uts{\tau}]$ and $\dot{\boldsymbol{v}}_{\tau-\sigma}$, and encode the resulting sequences using the CC encoder to produce the parity sequence $\dot{\boldsymbol{v}}_{\tau}$.
\end{enumerate}
The code rate of a HSC-BCC is $R=\frac{2L-\sigma}{6L-\sigma}\overset{\textrm{\scalebox{0.8}{$L\!\rightarrow\!\infty$}}}{=}1/3$. Fig. \ref{fig:hscpcc_enc} depicted the block diagram of an HSC-BCC encoder with $\sigma=2$, i.e., coupling memory $m=1$.

In contrast to most SC-TCs, HSC-BCCs use a simplified and different coupling pattern:
\begin{itemize}
\item \emph{Most SC-TCs:} The sequences $\boldsymbol{u}_t$ and/or $\boldsymbol{v}_t$ are decomposed into $m+1$ subsequences, and each component encoder takes $m$ coupling inputs from the preceding $m$ time instants.
\item \emph{HSC-BCCs:} The sequences $\uts{\tau}$ and $\vt{\tau}$ are directly used as the coupling inputs without further decomposition, and each component encoder takes only two coupling inputs from half-time instants $\tau+\sigma-1$ and $\tau+\sigma$.
\end{itemize}

This structure greatly simplifies the encoding and decoding processes. More importantly, HSC-BCCs are numerically observed to exhibit threshold saturation and attain very close-to-capacity performance for a wide range of rates of practical interest \cite[Table II]{xw_hscbcc}, with a coupling memory as small as 2 and under sliding window decoding.

\subsection{Spatially Coupled Coding Variants}
In addition to the prominent SC-LDPC codes and SC-TCs, a range of spatially coupled codes have been proposed in the literature for different communication scenarios or addressing some of the practical limitations of the conventional spatially coupled codes. In this subsection, we first review two representative classes of such codes: (i) staircase and zipper codes, and (ii) spatially coupled sparse regression codes (SC-SPARCs). We then review two variants of spatial coupling structures.

\vspace{0.5em}
\subsubsection{Staircase Codes and Zipper Codes}\label{sec:stair}

The channel coding schemes for high-speed optical transport networks to support a throughput approaching 1 Tb/s are required to achieve high coding gains at a BER below $10^{-15}$. Often, hard-decision decoding (HDD) is favored over soft-decision decoding due to its lower complexity and power consumption (per decoded bit). In this context, spatially coupled product-like codes with algebraic component codes, such as braided block codes \cite{4957627,6831429} and staircase codes \cite{Smith12} with BCH component codes, have been investigated for optical communication systems. Notably, \cite{Smith12} demonstrates that staircase codes can achieve a bit error rate (BER) of $10^{-15}$ within 0.5 dB of the BSC capacity under iterative HDD.

Staircase codes are naturally unterminated \cite{Smith12} and can be viewed as the result of applying spatial coupling to product codes. Following Fig. \ref{fig:staircase}(a), staircase codes are defined by a sequence of code blocks $\boldsymbol{X}_t$ indexed by $t\in \mathbb{N}$ and $\mathbb{N}$ denotes the set of natural numbers. Consider $m=1$. Each row of $[\boldsymbol{X}^{\mathsf{T}}_t,\boldsymbol{X}_{t+1}]$ is a valid codeword of an $(n,k)$ code $\mathcal{C}$, where $n$ is an even number. Initialize $\boldsymbol{X}_{0} = \boldsymbol{0}$. The encoding process at time $t>0$ can be described as follows:
\begin{enumerate}
\item Form the matrix $[\boldsymbol{X}^{\mathsf{T}}_{t-1},\boldsymbol{U}_t]$, where $\boldsymbol{U}_t$ contains new information bits.
\item Perform row-by-row encoding of the resulting matrix to obtain $[\boldsymbol{X}^{\mathsf{T}}_{t-1},\boldsymbol{U}_t,\boldsymbol{P}_t]$, where each row of $\boldsymbol{P}_t$ contains the corresponding parity bits.
\item Construct the code block as $\boldsymbol{X}_t = [\boldsymbol{U}_t, \boldsymbol{P}_t]$.
\end{enumerate}

The code structure is illustrated in Fig. \ref{fig:staircase}(a), where the horizontal and vertical codewords are highlighted. The size of each staircase code block $\boldsymbol{X}_i$ is $\frac{n}{2} \times \frac{n}{2}$. This gives the overall code rate of a staircase code as $\frac{2k}{n}-1$. As mentioned before, staircase codes with BCH component codes have been adopted as a part of the OIF 400 ZR \cite{400ZR}, ITU-T G.709.2 \cite{G709}, and IEEE 802.3ct standards \cite{IEEE802}.

\begin{figure}[t!]
	\centering
\includegraphics[width=\linewidth]{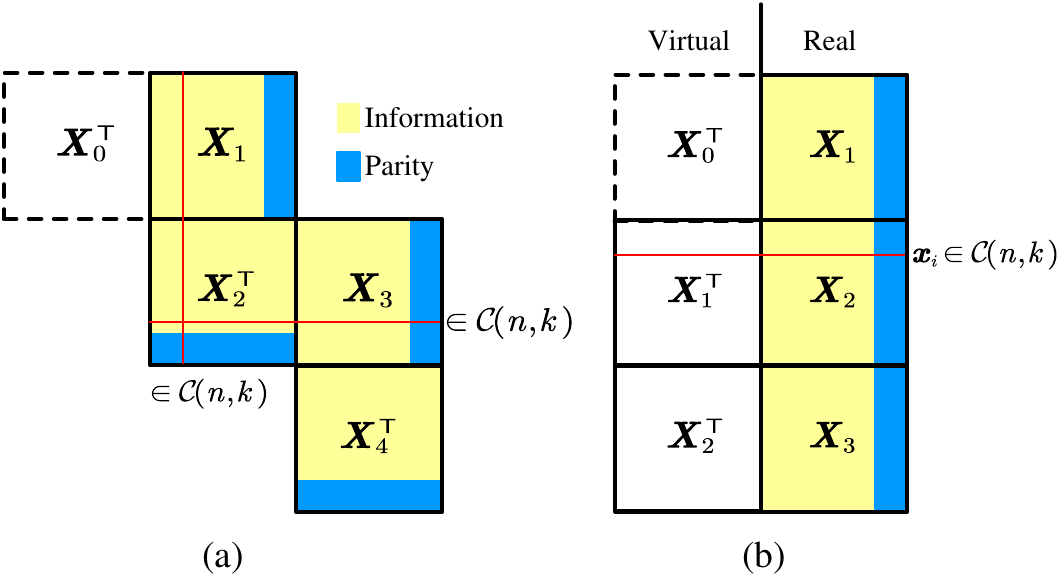}
\caption{Block diagram of (a) a staircase code, and (b) the corresponding zipper code representation.}
\label{fig:staircase}
\end{figure}

However, the staircase code block size (and encoder memory size) grows explosively as the code rate approaches 1 \cite[Fig. 11]{Sukmadji}. This is because the underlying BCH component codes must have larger information and codeword lengths while maintaining the same error-correction capability. Hence, new code families that are memory-efficient at high rates are called for.

To address this issue, the authors in \cite{9843869} introduced the zipper code framework, which describes spatially coupled product-like codes and subsumes many well-known codes, such as staircase codes and braided block codes. Moreover, the zipper code framework provides a large design space, where one can identify many memory-efficient coding schemes within the framework. For example, \cite[Table 2.1]{Sukmadji} shows that a subclass of zipper codes called tiled diagonal zipper codes achieves BER performance comparable to staircase codes at rates 0.96, 0.97, and 0.98, while the block size of a tiled diagonal zipper code is only half that of a staircase code. Further results show that carefully designed tiled diagonal zipper codes can achieve a gap to the BSC capacity within 0.5 dB at rates higher than 0.97 \cite[Table I]{9843869}.

A zipper code consists of real bits and virtual bits, as illustrated in Fig. \ref{fig:staircase}(b). The real bits are transmitted, whereas the virtual bits participate in the encoding process but are not transmitted. The mapping from virtual bits to real bits is specified by an interleaver. A formal definition of a zipper code is provided below.

\begin{framed}
A zipper code (unterminated) is composed of a sequence of binary linear constituent codewords $\boldsymbol{x}_0,\boldsymbol{x}_1,\ldots,$ where for $i\in \mathbb{N}$, $\boldsymbol{x}_i$ is a codeword of an $(n,k)$ component code $\mathcal{C}$. Let $m_0,m_1,\ldots$ be a sequence of integers satisfying $0 \leq m_i \leq k,\forall i\in \mathbb{N}$. A zipper code is specified by a zipping pair $(\mathcal{A},\mathcal{B})$ and a bijective interleaver map $\phi:\mathcal{A} \rightarrow \mathcal{B}$.

\medskip
\noindent\emph{Virtual set:}
\[
\mathcal{A} \triangleq \bigcup_{i\in\mathbb{N}} \{(i,j): j\in \{1,\ldots,m_i\}\}.
\]

\noindent\emph{Real set:}
\[
\mathcal{B} \triangleq \bigcup_{i\in \mathbb{N}} \{(i,j):j\in \{m_i+1,\ldots,n\}\}.
\]

\noindent\emph{Interleaver map:} Each index $(i,j)\in \mathcal{A}$ is mapped to a unique index $\phi(i,j) = (i',j') \in \mathcal{B}$ with $i' < i$.

\medskip
\noindent For $(i,j)\in\mathcal{A}$, the bit $x_{i,j}$ is not transmitted since it is a copy of $x_{\phi(i,j)}$ with $\phi(i,j)\in \mathcal{B}$, whereas bits with indices in $\mathcal{B}$ are transmitted. For initialization, set $x_{i',j'}=0$ for $i'<0$.
\end{framed}

\noindent In general, a zipper code is encoded in a serial manner. However, depending on the interleaver map, the encoding can be performed in parallel across multiple rows in some cases, e.g., staircase codes.

\medskip
\noindent\textbf{Example:} Fig. \ref{fig:staircase}(b) illustrates a zipper code representation of the staircase code shown in Fig.~\ref{fig:staircase}(a). In this representation, the interleaver mapping function is given by $\phi(\frac{n}{2}i+r,j)=(\frac{n}{2}(i-1)+j,\frac{n}{2}+r)$ for $r\in \{1,\ldots,\frac{n}{2}\}$. The virtual set (uncolored) and real set (colored), as defined above, are also labeled in the figure.

\medskip
More recently, higher-order staircase codes were introduced in \cite{10897312}, allowing each coded bit to be protected by an arbitrary number of component codewords rather than only two as in the conventional product-like codes. It was demonstrated that employing a single-error-correcting code as component codes suffices to reach a BER error floor below $10^{-15}$.

\begin{figure}[t!]
	\centering
\includegraphics[width=\linewidth]{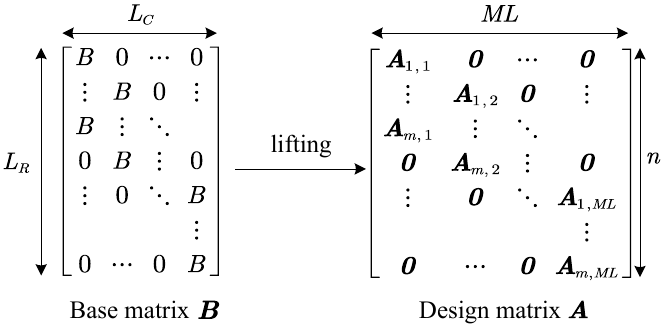}
\caption{Base matrix and design matrix of an SC-SPARC.}
\label{fig:SC_SPRACs}
\end{figure}

\vspace{0.5em}
\subsubsection{Spatially Coupled Sparse Regression Codes (SC-SPARCs)}
In addition to digital coding, spatial coupling has been applied to analog coding schemes. A popular class of such codes is sparse regression codes (SPARCs), introduced in \cite{6142076}, which are built on the ideas of compressed sensing. Unlike the aforementioned binary codes, e.g., SC-LDPC codes decoded using BP, SPARCs directly generate real-valued (or complex-valued) codewords and are decoded using approximate message passing (AMP). Notably, SPARCs unify coding and modulation in one step. In a standard SPARC, codewords of length $n$ are generated by $\boldsymbol{x} = \boldsymbol{A}\boldsymbol{\beta}$, where $\boldsymbol{A}\in \mathbb{R}^{n\times LM}$ is the design matrix and $\boldsymbol{\beta}\in \mathbb{R}^{LM}$ is a message vector that has exactly one non-zero entry in each of its $L$ sections. The power allocation across the nonzero entries of $\boldsymbol{\beta}$ has a crucial effect on its performance.

Spatial coupling has been applied to SPARCs to provide further performance enhancement. Similar to protograph SC-LDPC codes (see Sec. \ref{sec:scldpc}), the design matrix of an SC-SPARC is obtained from a base matrix via a procedure analogous to protograph lifting. Specifically, the design matrix $\boldsymbol{A}\in \mathbb{R}^{n\times LM}$, as shown in Fig. \ref{fig:SC_SPRACs}, is constructed from a base matrix $\boldsymbol{B}\in \mathbb{R}^{L_R \times L_C}$ by replacing each entry $B_{i,j}$ with an $\frac{n}{L_R} \times \frac{ML}{L_C}$ block $\boldsymbol{A}_{i,j}$ with i.i.d. entries $\sim\mathcal{N}(0,\frac{B_{i,j}}{L})$. Following Fig. \ref{fig:SC_gen_encoder}, the coupling of SC-SPARCs can be interpreted as feeding message vectors $\boldsymbol{\beta}_{t-m},\ldots,\boldsymbol{\beta}_{t}$ into the constituent SPARC encoder for $t\in \{1,\ldots,L_{\text{C}}\}$, where $\boldsymbol{\beta} = [\boldsymbol{\beta}_1,\ldots,\boldsymbol{\beta}_{L_{\text{C}}}]$. For initialization and termination, we set $\boldsymbol{\beta}_t=\boldsymbol{0}$ for $t<0$ and $t>L_{\text{C}}$, respectively. An SC-SPARC can then be decoded using a sliding-window AMP decoder, in a manner analogous to the sliding window BP decoder for SC-LDPC codes. In addition, the non-zero element in the base matrix is $B = P\cdot\frac{L_C+m-1}{m}$ to satisfy the power constraint $\frac{1}{L_RL_C}\sum^{L_R}_{i=1}\sum^{L_C}_{j=1}B_{i,j}=P$ and $m$ is the coupling memory. SC-SPARCs have a code rate of $R = \frac{L\log M}{n}$.

It was proven in \cite{8723607} that SC-SPARCs exhibit threshold saturation over general memoryless channels (GMCs), and analytical results therein further show that they universally achieve capacity. Moreover, SC-SPARCs provide superior finite-blocklength performance compared to uncoupled SPARCs. For example, SC-SPARCs with uniform power allocation are shown to achieve a rate gain exceeding 0.1 over uncoupled SPARCs with tailored power allocation at a codeword error probability below $10^{-3}$ for medium blocklengths \cite[Fig. 5.3]{CIT-092}. In fact, SC-SPARCs subsume standard SPARCs and power allocated SPARCs as special cases \cite[p42]{CIT-092}. As an example, the trivial base matrix with $L_{\text{R}}=L_{\text{C}}=1 $ corresponds to a standard uncoupled SPARC.

\begin{figure}[t!]
	\centering
\includegraphics[width=\linewidth]{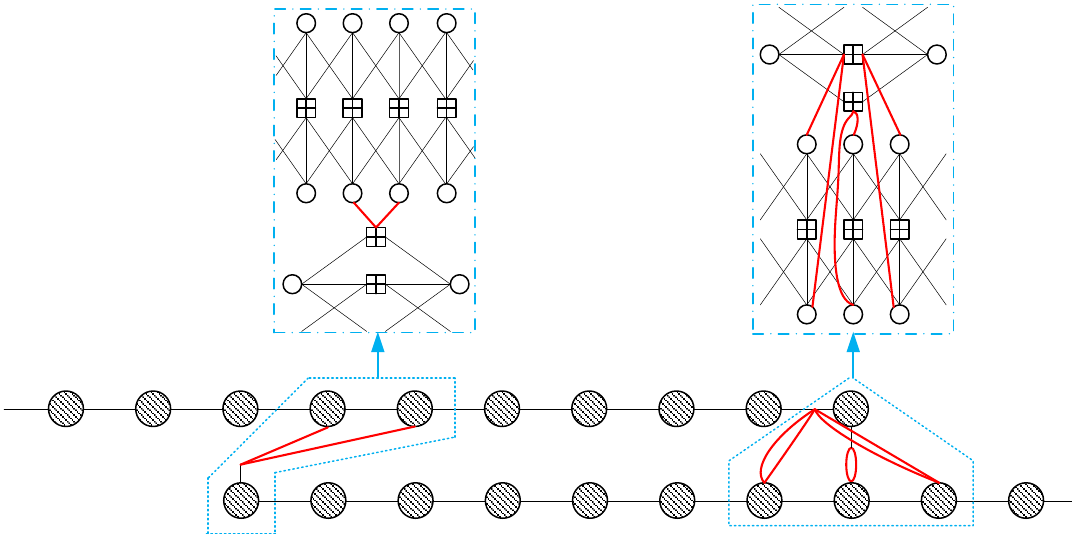}
\caption{Connected chain ensemble constructed from two $(3,6)$ single chain ensembles. Each shaded node illustrates a segment consisting of a CN and two VNs.}
\label{fig:multichain}
\end{figure}

\vspace{0.5em}
\subsubsection{Variants of Spatial Coupling Structures}

Beyond the choice of the base code, the coupling pattern itself can be designed to improve decoding thresholds, reduce rate loss, and enable local decodability.
In what follows, we review two recently proposed coupling structure designs. Although the methods below apply to SC-LDPC codes, they are also applicable to other spatially coupled codes. Due to space limitations, we only consider one-dimensional spatial coupling. For multi-dimensional coupling, the reader is referred to \cite{10449374}.

\vspace{0.5em}
\emph{3a) Connected Spatially Coupled Chains}

A connected chain ensemble is constructed by connecting several individual spatially coupled code chains \cite{8718022}. The main advantages of such constructions include:
\begin{itemize}
\item Multiple entry points for decoding waves, allowing reliable information to propagate from multiple directions;
\item Improved decoding thresholds, particularly at small coupling lengths;
\item A better trade-off between rate loss and gap to capacity.
\end{itemize}

To construct connected chain SC-LDPC code ensembles, additional edges are added to connect the termination CNs in one of the sub-chains to the VNs in other sub-chains. As an example, a connected chain ensemble constructed from two $(3,6)$ single chain ensembles is illustrated in Fig. \ref{fig:multichain}. The figure provides a simplified representation of the protograph, in which each shaded node denotes a segment consisting of one CN and two VNs. The connection point between the two chains was designed based on optimizing the decoding threshold, and one termination end of the two sub-chains is truncated to reduce the rate loss \cite{10172070}. It was shown that such an ensemble not only achieves a better trade-off between rate loss and gap to capacity than several existing protograph-based SC-LDPC codes with short chain lengths, but also exhibits threshold saturation behavior.

\vspace{0.5em}
\emph{3b) Spatially Coupled Codes with Sub-block Locality}

Applications such as data storage systems and low-latency data access benefit significantly from codes that support local decoding, as they enable fast access to small portions of data without decoding the entire codeword (i.e., global decoding). Motivated by this requirement, a new class of SC-LDPC codes with \emph{locally decodable} sub-blocks was proposed in \cite{9354799}, allowing either \emph{sub-block decoding} for fast access or full-block decoding for high data reliability.
\begin{framed}
\noindent\emph{Sub-block decoding} refers to a decoding method in which a target sub-block of a codeword is decoded by accessing the target sub-block and, optionally, a prescribed number of adjacent helper sub-blocks in the code chain.
\end{framed}

Sub-block decoding differs from sliding-window decoding in the following ways:
\begin{itemize}
\item Decoding can start close to the target sub-block, i.e., not necessarily at the first or last sub-block, allowing low-latency access to sub-blocks anywhere
in the block.
\item For a given target sub-block, there is no overlap between two window positions, which reduces latency and complexity.
\end{itemize}

Consider a $(d_{\text{v}},d_{\text{c}})$-regular SC-LDPC protograph constructed by coupling together a number of $(d_{\text{v}},d_{\text{c}})$-regular LDPC protographs with base matrix $\boldsymbol{B}_{\text{BC}} = \boldsymbol{1}^{d_{\text{v}} \times d_{\text{c}}}$. Let $T$ be the code's memory in units of the protograph size, such that $\boldsymbol{B}_{\text{BC}} = \sum^T_{\varsigma=0} \boldsymbol{B}_{\varsigma}$. To endow SC-LDPC codes with sub-block locality, a new coupling parameter $s\in\{1,\ldots,d_{\text{v}}-2\}$ was introduced in \cite{9354799} to designate the number of check nodes of a sub-block that connects to adjacent sub-blocks, in order to control the tradeoff between local and global decoding performance. Consider $T=1$. We define a protograph-based $(d_{\text{v}},d_{\text{c}},s)$-regular SC-LDPC code with sub-block locality as follows. Its protograph has essentially the same structure as that in Fig. \ref{fig:termination}(c) in Sec. \ref{sec:scldpc}, except that the component base matrix $\boldsymbol{B}_0$ contains $d_{\text{v}}-s \geq 2$ all-one rows and $s$ mixed rows containing both ones and zeros. The all-one rows correspond to \emph{local checks} that are connected only to VNs within individual sub-blocks, whereas the mixed rows correspond to \emph{coupling checks} that connect adjacent sub-blocks. The remaining component base matrix is given by $\boldsymbol{B}_1 = \boldsymbol{1}^{d_{\text{v}} \times d_{\text{c}}}-\boldsymbol{B}_0$. In other words, each sub-block is a $(d_{\text{v}}-s,d_{\text{c}})$-regular local code, and adjacent sub-blocks are connected via $s$ CNs whose connections are specified by the $s$ mixed rows in $\boldsymbol{B}_0$.

The parameter $s$ governs the trade-off between local and global decoding performance. A smaller value of $s$ implies that more CNs are local checks, resulting in improved local decoding performance at the expense of degraded global decoding performance. Conversely, for larger $s$, more CNs serve as coupling checks, yielding a more strongly coupled protograph, in which case the global decoding performance improves while the local decoding performance degrades. It should be noted that setting $T=1$ does not limit the code's global decoding performance, since the code's memory in units of VNs, i.e., coupling memory $m$ defined in Sec. II, can be increased by enlarging the dimension $d_{\text{c}}$ of base matrix $\boldsymbol{B}_{\text{BC}}$ with $T=1$ \cite[Remark 1]{9716111}. The connection point between sub-blocks can be optimized based on balancing the local and global decoding thresholds via DE.

\subsection{Discussion}

In this subsection, we summarize the similarities and differences among the spatially coupled codes introduced in this section. A comparison of these codes is summarized in Table \ref{tab:comparison1}.

\begin{table*}[t]
\centering
\caption{Comparison of spatially coupled (SC) code families}\label{tab:comparison1}
\begin{tabular}{|l|l|l|l|l|}
\hline
\textbf{SC Code Family} & \textbf{Component Codes} & \textbf{Construction} & \textbf{Decoder} & \textbf{Main Strengths} \\
\hline
SC-LDPC & LDPC codes & matrix-based & BP &
\begin{tabular}[c]{@{}l@{}}
capacity-achieving over general BMS channels, low error floor
\end{tabular} \\
\hline
GSC-PCCs & PCCs & \multirow{4}{*}{encoder-based} & \multirow{4}{*}{\begin{tabular}[c]{@{}l@{}} iterative \\ BCJR \end{tabular}} & \begin{tabular}[c]{@{}l@{}} BEC capacity-achieving, fully parallel encoding \end{tabular} \\
\cline{1-2} \cline{5-5}
SC-SCCs & SCCs &  &  & \begin{tabular}[c]{@{}l@{}} low error floor, fully parallel encoding \end{tabular} \\
\cline{1-2} \cline{5-5}
HSC-BCCs & BCCs &  &  & \begin{tabular}[c]{@{}l@{}} near-capacity BEC thresholds with a small coupling memory, \\ low error floor \end{tabular} \\
\hline
Staircase / Zipper & BCH product codes& encoder-based &
\begin{tabular}[c]{@{}l@{}}
iterative \\
HDD
\end{tabular} &
\begin{tabular}[c]{@{}l@{}}
BSC capacity-approaching at high rates, ultra-low error floor,\\ low-complexity high-throughput decoding
\end{tabular} \\
\hline
SC-SPARCs & SPARCs & matrix-based & AMP &
\begin{tabular}[c]{@{}l@{}}
capacity-achieving over GMCs, unified coding and modulation
\end{tabular} \\
\hline
\end{tabular}
\end{table*}

The spatially coupled codes defined through matrices, e.g., SC-LDPC codes and SC-SPARCs, have fixed structures and blocklengths once the corresponding matrices are specified. In contrast, the coupling of SC-TCs, staircase codes, and zipper codes is defined directly through their component code encoders. SC-TCs allow for flexible codeword lengths and structures, whereas staircase codes and zipper codes with algebraic component codes have fixed structures and codeword lengths. In addition, SC-PCCs and GSC-PCCs support fully parallel encoding, where $L$ component turbo encoders can operate simultaneously. This parallelism significantly reduces encoding latency compared to spatially coupled codes with serial encoding, e.g., HSC-BCCs.

It is also worth noting that the coupling pattern of HSC-BCCs differs from that of all other spatially coupled codes introduced in this section. Specifically, the coupling sequence from one time instant is re-encoded by the component encoder at a single different time instant. By comparison, for most other spatially coupled codes, the coupling sequences from one time instant are distributed across component encoders over multiple time instants. In addition, SC-SPARCs differ from most spatially coupled codes in the sense that they generate real-valued (or complex-valued) codewords, unifying coding and constellation shaping in one step. Furthermore, all spatially coupled codes considered in this section, except for SC-SPARCs, can be represented by a Tanner graph.

Spatially coupled codes exhibit several key trade-offs, as summarized below:
\begin{itemize}
\item \emph{Performance vs. decoding complexity:} In general, approaches such as increasing the coupling memory, employing additional repetitions, or adopting more complex coupling structures can further narrow the gap to capacity. However, these approaches increase the computational cost per decoding iteration, leading to higher decoding complexity.
\item \emph{Performance vs. decoding latency:} While a larger coupling memory improves decoding thresholds, it necessitates a proportional increased window size under sliding window decoding. As a result, the overall decoding latency is significantly increased.
\item \emph{Performance vs. rate loss:} A small coupling length can reduce the latency and memory cost of full-chain decoding, but often increases the gap to capacity due to a larger rate loss associated with termination.
\end{itemize}

From a design perspective, achieving a favorable balance between performance, complexity, and latency remains a central challenge. For many spatially coupled codes, such as regular SC-LDPC codes, GSC-PCCs, SC-SCCs, etc., achieving near-capacity performance typically requires a large coupling memory. Moreover, additional parameters often need to be increased accordingly, such as higher node degrees for regular SC-LDPC codes and larger repetition factors for GSC-PCCs. In contrast, HSC-SCCs can achieve thresholds close to capacity with a small coupling memory. Other coupling structures, such as the connected chain construction and sub-block locality structures, introduce new tradeoffs between decoding latency and performance.

\section{Useful Properties of Spatially Coupled Codes}\label{sec:3}

\subsection{Threshold Saturation}\label{sec:thres_sa}
Informally, spatial coupling enables BP decoding to achieve performance close to that of the optimal MAP decoding. For a spatially coupled ensemble $\mathcal{C}_{\text{s}}(m,L)$ with coupling memory $m$ and coupling length $L$ and its corresponding uncoupled counterpart $\mathcal{C}_{\text{u}}$, threshold saturation refers to the property that
\[
\lim\limits_{m \rightarrow \infty}\lim\limits_{L \rightarrow \infty}\theta_{\text{BP}}(\mathcal{C}_{\text{s}}) \rightarrow  \theta_{\text{MAP}}(\mathcal{C}_{\text{u}}),
\]
where $\theta$ denotes the decoding threshold for a general memoryless channel, such as the BEC, BSC, or additive white Gaussian noise (AWGN) channel.

Randomized SC-LDPC code ensembles have been proven to have threshold saturation over the BEC \cite{5571910,5695130} and the general BMS channel \cite{6912949}. Meanwhile, a simplified and general framework to prove threshold saturation for a broad class of coupled systems was proposed in \cite{6325197}. This simple technique has been used to prove threshold saturation for SC-LDPC codes and several classes of SC-TCs, e.g., \cite{Moloudi-scTurbo,min_gscpcc}. The key ingredient is the potential function.

The \emph{potential function} is a scalar functional of the DE state whose stationary points have a one-to-one correspondence with the fixed points of the DE recursion corresponding to the uncoupled system, and whose local minima correspond to stable fixed points. The potential threshold of the uncoupled system can be conveniently computed using the potential function. It is defined as the largest value of the channel parameter below which the potential function has a unique global minimum at the zero fixed point. For channel parameters between the BP and potential thresholds, the potential function of the uncoupled system admits multiple fixed points. In this regime, the DE recursion leads to a non-zero fixed point associated with BP decoding failure. The corresponding coupled system, on the other hand, initiates a decoding wave that propagates from each boundary toward the center of the chain, allowing its BP threshold to surpass that of the uncoupled system.

It was proven in \cite[Th. 1]{6325197} that for channel parameter below the potential threshold, the coupled recursion corresponding to the spatially coupled system has a unique fixed point that corresponds to the minimum of the potential function. Moreover, by combining \cite[Lemma 49]{6887298} and \cite[Sec. II-B]{7426827}, one can analytically show that the potential threshold is equal to the MAP threshold for a large class of uncoupled LDPC code ensembles. For uncoupled turbo-like code ensembles such as PCCs and SCCs, numerical evidence is provided in \cite{8002601} that the potential threshold matches the MAP threshold. For SC-GLDPC codes under iterative hard-decision decoding \cite{7954697}, the decoding threshold saturates instead to an intrinsic threshold defined by the suboptimal component decoders. Nevertheless, the implication is that spatial coupling allows the underlying codes to achieve the MAP decoding performance with BP decoding.

The potential function argument was later generalized to establish threshold saturation for randomized non-binary SC-LDPC codes on the BEC \cite{7430313}, as well as for randomized SC-LDPC, spatially coupled low-density generator-matrix (LDGM) codes \cite{6912949} on BMS channels, and SC-SPARCs \cite{hsieh_2021} on Gaussian channels.

For some structured spatially coupled codes, such as protograph-based SC-LDPC codes in Sec. \ref{sec:scldpc} and HSC-BCCs in Sec. \ref{sec:sctc}, threshold saturation has been numerically observed but not analytically proven. In these ensembles, the coupling structure is deterministic rather than random, which prevents a direct application of existing analytical techniques developed for randomly coupled systems. Nevertheless, it was shown in \cite{7541672} that there exists a family of deterministic braided codes that follows the same DE recursion as the corresponding ensemble, and hence threshold saturation can be established. However, the result is obtained in the high-rate regime (i.e., with code rate approaching 1). Establishing a general and rigorous proof of threshold saturation for structured spatially coupled codes therefore remains an open problem; see Sec.~\ref{sec:str_sc_thr} for further discussion.

\subsection{Universality}
Conventional uncoupled block codes, such as irregular LDPC codes designed for one specific channel, may not perform well on other channels. For example, an irregular LDPC code ensemble optimized to approach the BEC capacity typically requires a substantially different degree distribution to achieve comparable performance on the AWGN channel. In contrast, spatially coupled codes exhibit a form of \emph{universality}, meaning that a single ensemble works well for many channels, without requiring channel-specific optimization (assuming that the decoder has the channel knowledge).

In \cite{6589171}, it was analytically proven that for any $\delta\in(0,1)$, there exists a regular SC-LDPC code ensemble that achieves at least a fraction $1-\delta$ of the BMS channel capacity universally under BP decoding, with vanishing block error probability. The key insight is that regular SC-LDPC code ensembles achieve the \emph{area threshold} of the underlying uncoupled ensembles. For regular LDPC code ensembles, the area threshold upper bounds the MAP threshold, and both approach channel capacity as the degrees become large.

The area threshold is a functional of the generalized extrinsic information transfer curve associated with BP decoding (BP-GEXIT) \cite[Eq. (9)]{7426827} and is defined as the channel parameter value at which the area under this curve, measured from that value up to 1 (corresponding to a useless channel), equals the design rate. Moreover, a BP-GEXIT chart characterizes the asymptotic behavior of BP decoding by measuring the derivative of the conditional entropy of the transmitted codeword with respect to a channel parameter, conditioned on the channel output and the extrinsic information generated by BP decoding.

The universality of spatially coupled codes has been extended beyond the BMS channel. In \cite{6120387}, it was empirically shown that regular SC-LDPC code ensembles universally approach the capacity of the two-user binary-input Gaussian multiple access channel. Even though the channel gains are assumed to be unknown at the transmitter, reliable communication is achievable by employing SC-LDPC codes and joint BP decoding, for almost all channel gains that are theoretically possible. The universality of regular SC-LDPC codes over intersymbol interference (ISI) channels has been supported by numerical evidence in \cite{9611454}. In essence, the universality of spatially coupled codes is established based on generalizing threshold saturation over different channels.

\subsection{Linear Minimum Distance Growth}
Minimum distance is a key indicator of error floor performance under maximum-likelihood (ML) decoding. Moreover, linear minimum distance growth guarantees that as the blocklength tends to infinity, the codes do not suffer from the error floor phenomenon under ML decoding. In addition, the error floor predicted from the minimum distance can serve as a lower bound on the error floor performance under suboptimal iterative decoding.

Several classes of spatially coupled codes have been shown to have linear minimum distance growth. The analysis is often based on weight enumerators. For protograph-based $(d_{\text{v}},d_{\text{c}})$-regular SC-LDPC codes, it was shown that the minimum distance grows linearly with the component codeword length (i.e., lifting factor) \cite{Mitchell2015sc}. The same phenomenon has been observed for connected chain ensembles \cite{8718022} and some classes of SC-TCs, e.g., SC-BCCs \cite{Moloudi_sctc_distance}.

In fact, several works have shown that spatial coupling either improves or preserves the minimum distance of the underlying uncoupled codes under some mild conditions \cite{6262475,Moloudi_sctc_distance}. The implication is that the error floor performance of a spatially coupled code can be improved or analyzed by using the results from its underlying uncoupled code. In addition, recent work \cite{Battaglioni2024sc} shows that the free distance, which is known as the convolutional counterpart of the minimum distance, can be further improved by designing periodically time-varying SC-LDPC codes with a small period of variability.

\subsection{Finite-Length Scaling Law}\label{sec:3c}
To adopt spatially coupled codes in practical settings, it would require specifying a number of parameters, including the underlying component codes, the length, the coupling structures, and the decoder configurations. Understanding the impacts of these parameters on the performance is of vital importance. While the asymptotic behavior of spatially coupled codes is well understood, their finite-length behavior in the waterfall region requires further study. In this subsection, we review the recent results on the finite-length scaling law for spatially coupled codes and explain the quantities of interest, their modeling, and the factors leading to finite-length decoding failures.

Non-asymptotic performance characterizations have been developed primarily for SC-LDPC codes under BP decoding on the BEC in the form of finite-length scaling laws \cite{7086074,10068559}. The key approach is to analyze the stochastic process associated with the \emph{fraction of degree-one CNs} in the residual graphs obtained during peeling decoding (PD). A decoding failure occurs when the peeling decoder runs out of degree-one CNs before recovering all VNs. Before proceeding, we first introduce the following key concepts used in the analysis of the finite-length scaling law.
\begin{itemize}
\item \emph{Fraction of degree-one CNs:} $r^{(\ell/N)}_1$, where $\ell$ is the PD iteration and $N$ is the codeword length of the component LDPC code.
\item \emph{Steady-state phase:} the regime where the mean $\mathbb{E}[r^{(\ell/N)}_1]$, averaged over the ensemble, channel, and peeling decoding realizations, remains essentially constant.
\item \emph{First hit time:} $\tau_0 = \min\{\ell/N:r^{(\ell/N)}_1=0\}$, the normalized time of PD at which the number of degree-one CNs and, and hence $r^{(\ell/N)}_1$, drops to zero.
\end{itemize}
\medskip

As observed in \cite{7086074}, the mean of the fraction of degree-one CNs $\mathbb{E}[r^{(\ell/N)}_1]$ remains essentially constant during the steady-state phase. Moreover, since $r^{(\ell/N)}_1$ drops to its lowest value in this phase, decoding failure is most likely to occur therein. With the first hit time $\tau_0$ defined earlier, the probability of decoding failure $P_{\text{f}}$ can then be obtained by estimating the probability that $\tau_0$ is within the steady state phase. Most importantly, $r^{(\ell/N)}_1$ in the steady state is modeled as an Ornstein–Uhlenbeck (OU) process \cite{7086074}. The associated distribution of $\tau_0$ is approximated by an exponential distribution whose mean is a function of $\mathbb{E}[r^{(\ell/N)}_1]$, variance $\text{Var}[r^{(\ell/N)}_1]$, and covariance $\text{Cov}[r^{(\ell/N)}_1,r^{(\ell'/N)}_1]$ with $\ell \neq \ell'$. Let $\epsilon$ and $\epsilon_{\text{BP}}$ denote the channel erasure probability and the BP threshold on the BEC, respectively. Consequently, the frame error rate $P_{\text{f}}$ for an SC-LDPC code ensemble with a given degree distribution can be expressed as a function of $N$, $L$, $\epsilon$, $\epsilon_{\text{BP}}$, and scaling parameters derived from the aforementioned first- and second-order moments of $r^{(\ell/N)}_1$.

The most recent result on finite-length scaling laws is due to \cite{10068559}, which currently provides the most accurate characterization for randomized regular SC-LDPC codes on the BEC under sliding-window decoding with a limited number of iterations. In particular, windowed decoding with a window size of $W$ was analyzed as a ``two-phase'' procedure, where the first phase comprises the $L-W$ positions and includes a single decoding wave from the left, and the second phase comprises the last $W$ positions and may contain decoding waves propagating from both the left and the right. A decoding wave stops propagating either because all degree-one CNs are exhausted or it reaches the maximum number of decoding iterations, leading to decoding failure. In the first decoding phase, $r^{(\ell/N)}_1$ is modeled as an OU process following \cite{7086074}, whereas in the second phase it is modeled as the sum of two independent OU processes for the left and right waves.

The limit on the number of decoding iterations also affects the probability of decoding errors in two ways \cite{10068559}. First, it introduces a race between the left-propagating decoding wave and the sliding window. Specifically, if the left boundary of the window overtakes the wave at any iteration, decoding fails, even though successful decoding might have been possible in the absence of an iteration limit. Let $\text{Pr}\{O\}$ denote the probability of this overtaking event. To estimate $\text{Pr}\{O\}$, the stochastic process governing the position of the left wave is modeled as a scaled time integral of an OU process. The evolution of the corresponding probability density function across iterations is then tracked via the Fokker–Planck equation, which is solved numerically.

Second, once the sliding window reaches the right boundary of the coupled code chain, the iteration limit reduces the maximum propagation distance of the right wave. To account for this iteration limit, we denote by $W'\leq W$ the \emph{reduced window size}, reflecting the reduced number of positions the right wave can propagate to the left as the sliding window moves to the right, based on estimating the average number of erased VNs in the middle of the coupled chain during the steady state. Note that $W'$ is used for analysis only and not for simulation. In light of the above discussion, the finite-length scaling law of SC-LDPC codes under windowed decoding with a limited number of iterations is given by \cite[Eq. (54)]{10068559},
\[
P_{\text{f}} = 1-(1-\text{Pr}\{O\})(1-P^{(L-W')}_{\text{f},1})(1-P^{(W')}_{\text{f},2}),
\]
where $P^{(L-W')}_{\text{f},1}$ denotes the decoding failure probability in the first decoding phase with one decoding wave, $P^{(W')}_{\text{f},2}$ denotes the decoding failure probability in the second decoding phase with two decoding waves. In words, failure occurs if the wave is overtaken or if decoding fails in either phase. Although the above scaling law is for SC-LDPC codes, it can still provide insights into the non-asymptotic behavior of other coupled codes.

\section{Design of Spatially Coupled Codes}\label{sec:4}
The optimization of spatially coupled codes can be performed from both construction and decoding. Representative examples include optimizing the code structure to improve error rate performance \cite{Mitchell2015sc,Moloudi-scTurbo,Moloudi_sctc_distance} and mitigating error propagation under sliding window decoding \cite{9165839,10243110}. In this article, we focus on the design of spatially coupled codes from three perspectives: decoding threshold optimization, error floor optimization, and finite-length optimization.

\subsection{Decoding Threshold Optimization}
Existing capacity-achieving spatially coupled codes require several parameters—such as node degrees, coupling width, and coupling length for a regular SC-LDPC code—to tend to infinity. In practice, however, these parameters must be finite, which inevitably results in a gap to capacity. To reduce the gap, it is necessary to optimize the decoding threshold by using some analytical tools such as DE and potential functions \cite{6325197}.

For spatially coupled codes with the threshold saturation property, the optimization can be simplified to designing coding parameters to optimize the MAP threshold of the underlying uncoupled codes using the potential function. In most cases, when the coupling width and coupling length are moderately large, the BP threshold of the coupled ensembles and the MAP threshold of the corresponding uncoupled ensembles become identical to several decimal places. This justifies optimizing the MAP threshold as a practical method for designing spatially coupled codes. For example, for SC-SCCs on the BEC, the puncturing pattern that yields the largest BP threshold is to fully puncture the parity bits of the outer component encoder. This choice follows directly from maximizing the MAP threshold of the uncoupled SCCs \cite{Moloudi-scTurbo}.

When the coupling memory or coupling length or both are required to be small due to some complexity constraint, e.g., low-latency windowed decoding, optimizing the MAP threshold of the uncoupled codes may not be effective as directly optimizing the iterative decoding threshold of the corresponding coupled codes using DE. For the latter case, one can perform the DE optimization on the BEC and approximate the results to other channels, e.g., AWGN channel, to reduce optimization complexity \cite{Moloudi_sctc_distance}.

In summary, for spatially coupled codes with moderately large coupling memory and coupling length, decoding threshold optimization can be effectively achieved by optimizing the MAP threshold of the underlying uncoupled codes, provided that the codes exhibit threshold saturation. In contrast, for small coupling parameters, it is preferable to directly optimize the iterative decoding threshold using DE on the BEC.

\subsection{Error Floor Optimization}

For spatially coupled graph-based codes, the error floor performance is adversely affected by the existence of cycles in their code graphs. Therefore, maximizing the minimum cycle length, known as the girth of the code graph, is always of interest in code design.

To improve the error floor performance of SC-LDPC codes, recent works \cite{9112247} and \cite{9354189} introduced periodically time-varying SC-LDPC codes to achieve a large girth.
Compared to the time-invariant counterparts, these codes have a small period of variability in their corresponding code graphs after the lifting procedure.
As such, the degrees of freedom are increased during the code construction so that there will be potentially more space to optimize for large girths.
Applying combinatorial objects to code design also offers a promising approach to controlling the girth. For example, the difference triangle sets are adopted in \cite{10897312} to create an infinite 4-cycle-free code graph. Similar methods are also utilized in \cite{1053951,Alfarano2020ConstructionOL,6510023} to achieve a low error floor via enlarging the girth of the constructed code graph.

Another important factor affecting the error floor is the coupling memory. For both SC-LDPC codes and SC-TCs, increasing the coupling width has been shown to lower the error floor in some cases. For example, \cite{9112247} shows that increasing the coupling width makes it easier to eliminate 4-cycles in protograph-based quasi-cyclic SC-LDPC codes. When combined with their proposed lifting design, this can yield a lower error floor compared with constructions that use a smaller coupling width. For SC-SCCs, \cite{10197554} shows that the error floor improves as the coupling memory increases via simulation.

Further lowering the error floor requires a joint design of permutation, puncturing, and coupling structures.
As an example, the authors in \cite{sctc_convPerm} proposed a set of periodic convolutional permuters for SC-SCCs, where each permuter is obtained by unwrapping a dedicated block permuter to preserve the spread. Numerical results show that SC-SCCs with the designed periodic convolutional permuters have no visible error floor even when the information length of the component SCC is as small as 64.

The choice of component codes also influences the error floor. For GSC-PCCs, it was shown that using a state-8 convolutional component code achieves a lower error floor than that with a state-4 convolutional component code \cite{min_gscpcc}. Another example is staircase codes and their variants, where employing component BCH codes with stronger error-correction capability can significantly reduce the error floor, at the cost of a lower code rate.

In summary, for spatially coupled graph-based codes, reducing the error floor relies primarily on increasing the girth through code construction. For both SC-LDPC codes and SC-TCs, increasing the coupling memory can lower the error floor. Additional improvements can be achieved through joint optimization of permutations, puncturing, and coupling structures. Finally, the use of stronger component codes can also contribute to lowering the error floor.

\subsection{Finite-Length Optimization}
Recently, the finite-length scaling law has been utilized to optimize the waterfall performance of short-to-medium length spatially coupled codes under sliding window decoding with a target window size \cite{9732353}. Motivated by the finite-length scaling law, e.g., \cite{7086074}, \cite{9732353} proposed a new metric to reflect the finite-length performance of protograph-based SC-LDPC codes on the BEC under sliding window decoding. Rather than designing SC-LDPC codes directly based on the full scaling law, the codes were optimized using this metric, thereby reducing the optimization complexity. For clarity, the key aspects are summarized as follows:
\begin{itemize}
\item \emph{Optimization target:} improve the finite-length (waterfall) performance under windowed decoding for a given window size.
\item \emph{Surrogate metric:} window mean parameter, defined as the mean of the normalized fraction of degree-one CNs, where the normalization is with respect to the gap between the channel erasure probability and the windowed decoding threshold.
\item \emph{Practical benefits:} improved finite-length performance and reduced optimization complexity compared to directly optimizing the windowed decoding threshold.
\end{itemize}

As shown in \cite{7086074} (see also Sec. \ref{sec:3c}), the key to analyzing the finite-length scaling behavior of SC-LDPC codes under BP decoding on the BEC lies in the evolution of the number of degree-one CNs in the residual graph during the PD process, as its statistical properties depend on the blocklength as well as other coding parameters. Define the fraction of degree-one CNs as the number of degree-one CNs across all time instants, normalized by the codeword length of the component LDPC codes. The mean of this fraction is approximated by the change in the average erasure probability of VNs across all time instants between two consecutive windowed decoding iterations \cite[Alg. 1]{9732353}. In the context of BP decoding, this mean value is equal to the mean BP parameter multiplied by the gap between the channel erasure probability and the BP threshold \cite[Eq. (18)]{7086074}.

Following this formulation, the authors in \cite{9732353} introduced a window mean parameter, analogous to the mean BP parameter, as defined in the summary above. To this end, \cite{9732353} designed non-uniform coupling widths for protograph-based SC-LDPC codes based on optimizing the window mean parameter. Although accurately predicting the finite-length performance requires several scaling parameters and can be computationally demanding, it was empirically observed that the protograph-based SC-LDPC codes based on optimizing the window mean parameter achieve better finite-length performance than those based on optimizing the windowed decoding threshold alone for given target window sizes. Interestingly, \cite{9732353} also reports that the optimization time was relatively shorter than maximizing the windowed decoding threshold since their proposed method only runs the DE equations at a particular point.

\begin{table}[t]
	\caption{Configurations and BP thresholds over the BEC} \label{tb:sctc_de}
	\setlength{\tabcolsep}{4pt}
	{\scalebox{1}{
			\begin{tabular}{lp{45pt}cccc}
				\toprule
				&  & \multicolumn{2}{c}{$\ebp^{(R=1/2)}$} & \multicolumn{2}{c}{$\ebp^{(R=3/4)}$} \\
				\cmidrule[\lightrulewidth](l{2pt}r{2pt}){3-4}  \cmidrule[\lightrulewidth](l{2pt}r{2pt}){5-6}
				Ensemble & Component & $m=1$ & $m=2$ & $m=1$ & $m=2$\\
				\midrule
				\multirow{2}{*}{$\begin{matrix}
						(3,6) \hfill \\ \text{SC-LDPC}
					\end{matrix}$} & \scalebox{0.85}{$\begin{matrix}
						\boldsymbol{B}_0 = [2 2] \\ \boldsymbol{B}_1 = [1 1]
					\end{matrix}$} & 0.4881 & - & 0.2321 & - \\
				\cmidrule[\lightrulewidth](l{2pt}r{2pt}){2-6}
				& \scalebox{0.85}{$\begin{matrix}
						\boldsymbol{B}_i = [1 1] \\ \forall i = 0,1,2
					\end{matrix}$} & - & 0.4881 & - & 0.2321\\
				\midrule
				GSC-PCC-a & $[1,5/7]$ & 0.4883 & 0.4936 & 0.2240 & 0.2351  \\
				\cmidrule[\lightrulewidth](l{2pt}r{2pt}){2-6}
				GSC-PCC-b & $[1,15/13]$ & 0.4934 & 0.4965 & 0.2378 & 0.2428   \\
				\midrule
				SC-SCC & $[1,5/7]$ & 0.4789 & 0.4928 & 0.2230 & 0.2383  \\
				\midrule
				HSC-BCC &  ${\scalebox{0.9}{$\begin{bmatrix} 1, 0, 5/7 \\  0, 1, 3/7 \end{bmatrix}$}} $
				& 0.4993 & 0.4993 & 0.2497 & 0.2498 \\  %
				\midrule
				\multicolumn{6}{l}{Other configurations: }\\
				\multicolumn{6}{p{0.9\linewidth}}{$\ast$ For SC-TCs, the component generator polynomials are given in octal representation, e.g., $[15]_{\text{oct}}=1+x+x^3$.}\\	
				\multicolumn{6}{l}{$\ast$ GSC-PCCs: $q\!=\!2$; $\lambdar\!=\!1/3$ for $m\!=\!1$; $\lambdar\!=\!1/2$ for $m\!=\!2$.}\\
				\multicolumn{6}{p{0.9\linewidth}}{$\ast$ For GSC-PCCs and HSC-BCCs, parity sequences are punctured. For SC-SCCs, outer and inner parity sequences are punctured.}\\
				\bottomrule
	\end{tabular}}}
\end{table}

\section{Numerical Results}\label{sec:5}

In this section, we present numerical results for various spatially coupled codes, including BP thresholds over the BEC, waterfall performance at moderate blocklengths, and error floor performance at short blocklengths.

We consider protograph-based $(3,6)$-regular SC-LDPC codes and the three representative SC-TCs reviewed in Sec. \ref{sec:sctc}.  The following code parameters are common to both the DE results and the finite-length simulation results. Full BP decoding is assumed for DE, while the decoder configurations for finite-length simulations are presented later.
\begin{itemize}
\item \emph{Coupling length:} $L=50$, to reduce the rate loss due to termination.
\item \emph{Coupling memory:} $m\in\{1,2\}$, to limit encoding and decoding complexity.
\item \emph{Target rates:} $R\in\{1/2,3/4\}$, representing medium- and high-rate configurations.
\item \emph{Puncturing:} All codes are randomly punctured to achieve rates above their respective design rates.
\end{itemize}

The BP threshold for each spatially coupled code at a target rate $R$, denoted by $\epsilon^{(R)}_{\text{BP}}$, is reported in Table \ref{tb:sctc_de}, where GSC-PCC-a and GSC-PCC-b denote two configurations of GSC-PCCs. The remaining code-specific configurations are also provided in the same table. Notably, HSC-BCCs have the largest BP threshold and the smallest gap to capacity while requiring only a coupling memory of $m=2$. The thresholds of the other SC-TCs are comparable to those of SC-LDPC codes and improve with increasing coupling memory or with the use of stronger convolutional component codes.

\begin{figure}[t]
    \centering
    \includegraphics[width=\linewidth]{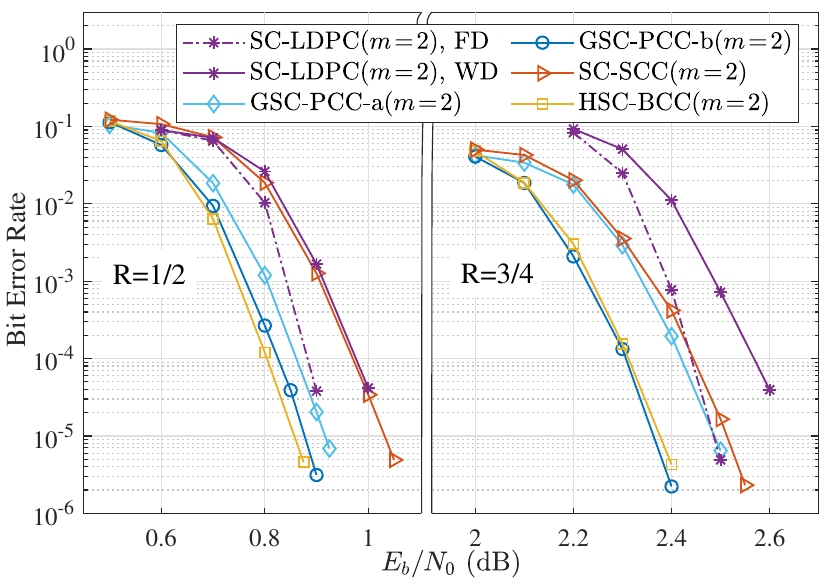}
    \caption{Waterfall performance of SC-LDPC codes with $M=2000$ and several SC-TCs with $K=2000$. ``FD'' denotes full BP decoding and ``WD'' denotes sliding window decoding. All SC-TCs are under sliding window decoding.}
    \label{fig:BER_K2000L50}
\end{figure}

\begin{figure}[t]
    \centering
    \includegraphics[width=\linewidth]{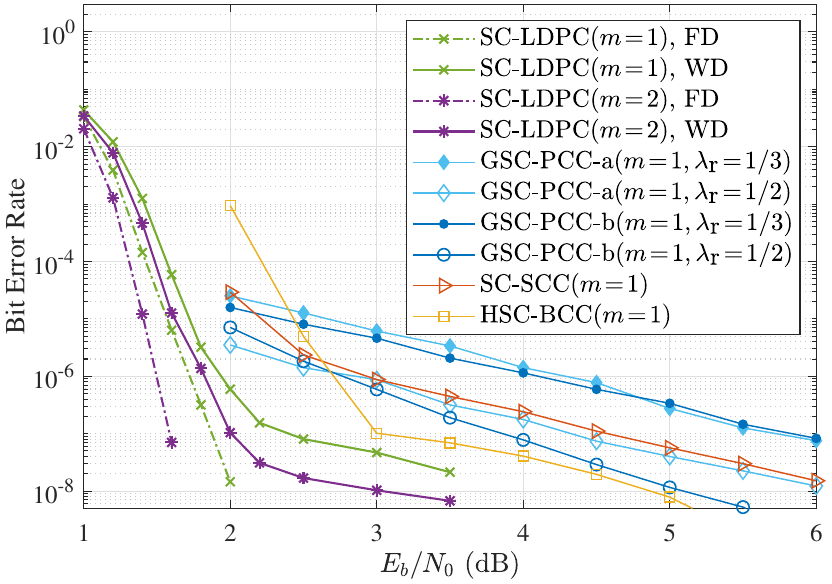}
    \caption{Error floor performance of SC-LDPC codes with $M=200$ and several SC-TCs with $K=200$. The BER results of SC-TCs with $m=2$ are omitted due to negligible error floor improvement compared with the case of $m=1$.}
    \label{fig:BER_errorFloor_K200L50}
\end{figure}

The BER performance for moderate and short blocklengths is shown in Figs. \ref{fig:BER_K2000L50} and \ref{fig:BER_errorFloor_K200L50}, respectively. The decoder configurations used in the simulations are summarized as follows:
\begin{itemize}
\item \emph{SC-LDPC:} full BP decoding (maximum 500 iterations); sliding window decoding (window size 12, maximum 200 iterations per window).
\item \emph{SC-TCs:} sliding window decoding (window size 8, maximum 20 iterations per window) with one inner decoding iteration for the base code.
\end{itemize}
For sliding window decoding, the window size and decoding iterations are chosen such that further increases in either parameter do not yield significant performance improvements.

As shown in Fig. \ref{fig:BER_K2000L50}, the codes with larger BP thresholds in Table \ref{tb:sctc_de} also achieve better waterfall performance under sliding window decoding for moderate blocklengths. In Fig. \ref{fig:BER_errorFloor_K200L50}, it is shown that SC-LDPC codes outperform all SC-TCs at short blocklengths. An exception is the HSC-BCC, whose error floor performance under sliding window decoding is comparable to that of SC-LDPC codes. Moreover, SC-LDPC codes under full-chain decoding do not show a visible error floor even with a small component codeword length.

In summary, the above results show that spatially coupled codes with larger BP thresholds over the BEC generally exhibit better waterfall performance over the AWGN channel at moderate blocklengths. This implies that spatially coupled codes exhibit a certain degree of universality when the blocklength is not small. However, good waterfall performance at short blocklengths and low error floors are not guaranteed, and require careful code design and appropriate decoder configurations.

\section{Open Problems and Future Directions}\label{sec:6}
In light of the work presented above, we outline the following open problems and potential future works of spatially coupled codes:
\begin{itemize}
\item Threshold saturation for structured coupled ensembles
\item Universality in non-asymptotic regimes and/or beyond BMS channels
\item High-throughput spatially coupled codes
\item Rate loss mitigation
\item Decoding wave propagation speed
\item Quantum error correction
\item Other applications of spatial coupling
\end{itemize}

\subsection{Open Problems}\label{sec:str_sc_thr}
\textbf{Open Problem 1}: Can threshold saturation be rigorously proven for spatially coupled ensembles with structured (deterministic) coupling?

So far, threshold saturation has only been analytically proven for randomized coupled ensembles. However, the fundamental question of whether the proof can be rigorously extended to coupled ensembles with certain structures remains open. Although the potential function argument \cite{6325197} provides a simple way towards a general threshold saturation proof, it does not provide an answer to the question of whether there exists a simpler coupled system to achieve threshold saturation. The work in \cite{xw_hscbcc} introduced a simple and deterministic coupling structure such that the coupled bits are delayed and encoded by the component encoder at one time instant. This is in sharp contrast to most coupled codes that spread the coupling bits across component encoders at multiple time instants. Notably, threshold saturation can be numerically observed with a coupling memory as small as 2. Furthermore, threshold saturation has been proven for a specific class of deterministic braided codes in the high-rate regime \cite{7541672}. Addressing the above fundamental question can provide guidance on designing powerful spatially coupled codes with theoretical performance guarantees and simple implementations.

\medskip
\textbf{Open Problem 2:} Can the universality of spatially coupled codes be retained in non-asymptotic regimes and/or extended beyond BMS channels?

Several works have shown that spatially coupled codes are universal for a range of channels, including general BMS channels \cite{6589171}, Gaussian multiple access channels \cite{6120387}, and ISI channels \cite{9611454}. Yet, the results were established based on the asymptotic setting, in the sense that many coding parameters need to approach infinity. In practical communication scenarios, the number of channel uses, decoding latency, and computational complexity are limited. In this case, it would be interesting to investigate how to maintain universality in non-asymptotic settings. One possible starting point can be generalizing the finite-length scaling law for spatially coupled codes \cite{10068559} to the general BMS channel and optimizing the coding parameters to achieve better finite-length performance, similar to the strategy used in \cite{9732353}. In addition, establishing the universality of spatially coupled codes for other channel models, such as general multiuser channels and channels with memory, is also an important direction for future research.

\subsection{Other Related Topics}
In addition to the open problems discussed above, several practical and application-oriented issues related to spatially coupled codes are worthy of further investigation.

\emph{High-throughput spatially coupled codes:} The 6G radio access network will require a peak throughput of 1 Tb/s. Channel coding is a key physical layer technique to support such a high data rate. Existing works have demonstrated sliding window decoders achieving throughput exceeding 100 Gb/s, e.g., \cite{10802915}. However, they still fall short of meeting the aforementioned throughput target. To facilitate the adoption of spatially coupled codes in future communication standards, it is imperative to investigate high-throughput decoding architectures and new coupling structures that enable a higher degree of parallelism.

\emph{Rate loss mitigation:} Directly terminating the coupled chain will inevitably introduce a rate loss. The rate loss can be significant and poses a bottleneck for very high-throughput applications. For SC-LDPC codes, the rate loss can be mitigated, for example, by using tail-biting, albeit at the cost of some performance degradation, or by connecting additional variable nodes to the boundary check nodes and optimizing them to avoid performance loss, e.g., \cite{8894067}. It remains important to develop systematic and effective approaches for reducing the rate loss of general spatially coupled codes.

\emph{Decoding wave propagation speed:} When decoding is triggered from the terminated boundaries, reliable information propagates along the spatially coupled code chain in a wave-like manner under iterative decoding. The speed of the decoding wave propagation plays a critical role in determining the decoding convergence speed, and consequently, the decoding complexity. Previous work, such as \cite{8419715}, has provided analytical bounds on the wave speed for SC-LDPC codes over general BMS channels. Understanding and optimizing the wave speed under practical settings, such as finite blocklength regimes and more general channels, is important for designing spatially coupled codes with strong practical performance.

\emph{Quantum error correction:} Quantum error correction is essential for fault-tolerant quantum communication and computation. The excellent performance of spatially coupled codes in classical communication, together with their compatibility with low-latency windowed decoding, has motivated interest in their quantum counterparts. For example, \cite{Yang2025spatiallycoupled} introduced an algebraic framework for spatially coupled quantum LDPC (SC-QLDPC) codes and used it to construct two-dimensional spatially coupled hypergraph-product codes with small memory and improved thresholds. Nevertheless, several important challenges remain to be addressed, including achieving higher code rates while maintaining favorable distance properties and mitigating decoding bottlenecks caused by short cycles.

\emph{Other applications:} The benefits of spatial coupling can be further explored for a wide range of applications. For example, the coupling structures can be leveraged to reduce the number of retransmitted blocks for hybrid automatic repeat request (HARQ). Spatial coupling can also be leveraged for multiple access, e.g., \cite{6171810,8638808}, where the coupling transmission structure can allow the suboptimal iterative receiver to attain the optimal MAP estimation performance. These observations motivate further investigation into the use of spatial coupling beyond channel coding.

\section*{Acknowledgment}

The authors would like to thank the Guest Editor and the anonymous reviewers for their valuable comments and insightful suggestions.

The work of Min Qiu was supported in part by the SJTU-ExploreX Funding under Grant SD6040004/153.
The work of Peng Kang was supported in part by the National Natural Science Foundation of China under Grant 62401150.
The work of Xiaowei Wu and Lei Yang was supported in part by the National Natural Science Foundation of China under Grant 62301533, 62371439, and U24A20214; in part by the National Key R$\&$D Program of China under Grant 2025YFF0522600.

\bibliographystyle{IEEEtran}
\bibliography{ref.bib}

\begin{thebibliography}{10}
\providecommand{\url}[1]{#1}
\csname url@samestyle\endcsname
\providecommand{\newblock}{\relax}
\providecommand{\bibinfo}[2]{#2}
\providecommand{\BIBentrySTDinterwordspacing}{\spaceskip=0pt\relax}
\providecommand{\BIBentryALTinterwordstretchfactor}{4}
\providecommand{\BIBentryALTinterwordspacing}{\spaceskip=\fontdimen2\font plus
\BIBentryALTinterwordstretchfactor\fontdimen3\font minus
  \fontdimen4\font\relax}
\providecommand{\BIBforeignlanguage}[2]{{%
\expandafter\ifx\csname l@#1\endcsname\relax
\typeout{** WARNING: IEEEtran.bst: No hyphenation pattern has been}%
\typeout{** loaded for the language `#1'. Using the pattern for}%
\typeout{** the default language instead.}%
\else
\language=\csname l@#1\endcsname
\fi
#2}}
\providecommand{\BIBdecl}{\relax}
\BIBdecl

\bibitem{782171}
A.~{Jimenez Felstrom} and K.~S. {Zigangirov}, ``Time-varying periodic
  convolutional codes with low-density parity-check matrix,'' \emph{IEEE Trans.
  Inf. Theory}, vol.~45, no.~6, pp. 2181--2191, Sep. 1999.

\bibitem{5571910}
M.~Lentmaier, A.~Sridharan, D.~J. Costello, and K.~S. Zigangirov, ``Iterative
  decoding threshold analysis for {LDPC} convolutional codes,'' \emph{IEEE
  Trans. Inf. Theory}, vol.~56, no.~10, pp. 5274--5289, Oct. 2010.

\bibitem{5695130}
S.~Kudekar, T.~J. Richardson, and R.~L. Urbanke, ``Threshold saturation via
  spatial coupling: Why convolutional {LDPC} ensembles perform so well over the
  {BEC},'' \emph{IEEE Trans. Inf. Theory}, vol.~57, no.~2, pp. 803--834, Feb.
  2011.

\bibitem{Smith12}
B.~P. Smith, A.~Farhood, A.~Hunt, F.~R. Kschischang, and J.~Lodge, ``Staircase
  codes: {FEC} for 100 {G}b/s {OTN},'' \emph{J. Lightw. Technol.}, vol.~30,
  no.~1, pp. 110--117, Jan. 2012.

\bibitem{400ZR}
{Optical Internetworking Forum (OIF)}, \emph{{Implementation Agreement 400ZR}},
  {OIF-400ZR-02.0}, Jul. 2022.

\bibitem{6912949}
S.~Kumar, A.~J. Young, N.~Macris, and H.~D. Pfister, ``Threshold saturation for
  spatially coupled {LDPC} and {LDGM} codes on {BMS} channels,'' \emph{IEEE
  Trans. Inf. Theory}, vol.~60, no.~12, pp. 7389--7415, 2014.

\bibitem{6589171}
S.~{Kudekar}, T.~{Richardson}, and R.~L. {Urbanke}, ``Spatially coupled
  ensembles universally achieve capacity under belief propagation,'' \emph{IEEE
  Trans. Inf. Theory}, vol.~59, no.~12, pp. 7761--7813, Dec. 2013.

\bibitem{6086762}
A.~R. Iyengar, M.~Papaleo, P.~H. Siegel, J.~K. Wolf, A.~Vanelli-Coralli, and
  G.~E. Corazza, ``Windowed decoding of protograph-based {LDPC} convolutional
  codes over erasure channels,'' \emph{IEEE Trans. Inf. Theory}, vol.~58,
  no.~4, pp. 2303--2320, 2012.

\bibitem{Mitchell2015sc}
D.~G.~M. {Mitchell}, M.~{Lentmaier}, and D.~J. {Costello}, ``Spatially coupled
  {LDPC} codes constructed from protographs,'' \emph{IEEE Trans. Inf. Theory},
  vol.~61, no.~9, pp. 4866--4889, Sep. 2015.

\bibitem{9398939}
D.~G.~M. Mitchell, P.~M. Olmos, M.~Lentmaier, and D.~J. Costello, ``Spatially
  coupled generalized {LDPC} codes: Asymptotic analysis and finite length
  scaling,'' \emph{IEEE Trans. Inf. Theory}, vol.~67, no.~6, pp. 3708--3723,
  2021.

\bibitem{Moloudi-scTurbo}
S.~{Moloudi}, M.~{Lentmaier}, and A.~{Graell i Amat}, ``Spatially coupled
  turbo-like codes,'' \emph{{IEEE} Trans. Inf. Theory}, vol.~63, no.~10, pp.
  6199--6215, Oct 2017.

\bibitem{1055186}
L.~Bahl, J.~Cocke, F.~Jelinek, and J.~Raviv, ``Optimal decoding of linear codes
  for minimizing symbol error rate,'' \emph{IEEE Trans. Inf. Theory}, vol.~20,
  no.~2, pp. 284--287, Mar. 1974.

\bibitem{min_gscpcc}
M.~Qiu, X.~Wu, J.~Yuan, and A.~{Graell i Amat}, ``Generalized spatially-coupled
  parallel concatenated codes with partial repetition,'' \emph{{IEEE} Trans.
  Commun.}, vol.~70, no.~9, pp. 5771--5787, 2022.

\bibitem{divsalar1998coding}
D.~Divsalar, H.~Jin, and R.~J. McEliece, ``Coding theorems for ``turbo-like''
  codes,'' in \emph{Allerton Conf. Commun., Control, Comp.,}, vol.~36, 1998,
  pp. 201--210.

\bibitem{5361461}
W.~{Zhang}, M.~{Lentmaier}, K.~S. {Zigangirov}, and D.~J. {Costello}, ``Braided
  convolutional codes: A new class of turbo-like codes,'' \emph{IEEE Trans.
  Inf. Theory}, vol.~56, no.~1, pp. 316--331, Jan. 2010.

\bibitem{4957627}
A.~J. {Feltstrom}, D.~{Truhachev}, M.~{Lentmaier}, and K.~S. {Zigangirov},
  ``Braided block codes,'' \emph{IEEE Trans. Inf. Theory}, vol.~55, no.~6, pp.
  2640--2658, 2009.

\bibitem{xw_hscbcc}
X.~Wu, L.~Yang, M.~Qiu, C.~Han, and J.~Yuan, ``Half spatially coupled
  turbo-like codes,'' in \emph{IEEE Inf. Theory Workshop (ITW)}, 2025, pp.
  1--7.

\bibitem{6831429}
Y.-Y. Jian, H.~D. Pfister, K.~R. Narayanan, R.~Rao, and R.~Mazahreh,
  ``Iterative hard-decision decoding of braided {BCH} codes for high-speed
  optical communication,'' in \emph{Proc. IEEE Globecom}, 2013, pp. 2376--2381.

\bibitem{G709}
{International Telecommunication Union - Telecommunication Standardization
  Sector (ITU-T)}, \emph{{OTU4 long-reach interface}}, {G.709.2 / Y.1331.2},
  Jul. 2018.

\bibitem{IEEE802}
{Institute of Electrical and Electronics Engineers (IEEE)}, \emph{{Physical
  Layers and Management Parameters for 100 Gb/s Operation over DWDM Systems}},
  {IEEE Std 802.3ct}, Jun. 2021.

\bibitem{Sukmadji}
A.~Y. Sukmadji, ``Zipper codes: High-rate spatially-coupled codes with
  algebraic component codes,'' Master's thesis, Univ. Toronto, Dept. Elect. and
  Comput. Eng., Univ. Toronto, Toronto, ON, Canada,, 2020.

\bibitem{9843869}
A.~Y. Sukmadji, U.~Martínez-Peñas, and F.~R. Kschischang, ``Zipper codes,''
  \emph{J. Lightw. Technol.}, vol.~40, no.~19, pp. 6397--6407, 2022.

\bibitem{10897312}
M.~Shehadeh, F.~R. Kschischang, A.~Y. Sukmadji, and W.~Kingsford,
  ``Higher-order staircase codes,'' \emph{IEEE Trans. Inf. Theory}, vol.~71,
  no.~4, pp. 2517--2538, 2025.

\bibitem{6142076}
A.~Joseph and A.~R. Barron, ``Least squares superposition codes of moderate
  dictionary size are reliable at rates up to capacity,'' \emph{IEEE Trans.
  Inf. Theory}, vol.~58, no.~5, pp. 2541--2557, 2012.

\bibitem{8723607}
J.~Barbier, M.~Dia, and N.~Macris, ``Universal sparse superposition codes with
  spatial coupling and {GAMP} decoding,'' \emph{IEEE Trans. Inf. Theory},
  vol.~65, no.~9, pp. 5618--5642, 2019.

\bibitem{CIT-092}
\BIBentryALTinterwordspacing
R.~Venkataramanan, S.~Tatikonda, and A.~Barron, ``Sparse regression codes,''
  \emph{Foundations and Trends in Communications and Information Theory},
  vol.~15, no. 1-2, pp. 1--195, 2019. [Online]. Available:
  \url{http://dx.doi.org/10.1561/0100000092}
\BIBentrySTDinterwordspacing

\bibitem{10449374}
H.~Esfahanizadeh, L.~Tauz, and L.~Dolecek, ``Harnessing degrees of freedom of
  spatially coupled graph codes for agile data storage,'' \emph{IEEE BITS the
  Information Theory Magazine}, vol.~3, no.~3, pp. 50--63, 2023.

\bibitem{8718022}
D.~Truhachev, D.~G.~M. Mitchell, M.~Lentmaier, D.~J. Costello, and A.~Karami,
  ``Code design based on connecting spatially coupled graph chains,''
  \emph{IEEE Trans. Inf. Theory}, vol.~65, no.~9, pp. 5604--5617, 2019.

\bibitem{10172070}
Y.~Liao, M.~Qiu, and J.~Yuan, ``Self-connected spatially coupled {LDPC} codes
  with improved termination,'' \emph{IEEE Commun. Lett.}, vol.~27, no.~8, pp.
  1959--1963, 2023.

\bibitem{9354799}
E.~Ram and Y.~Cassuto, ``Spatially coupled {LDPC} codes with sub-block
  locality,'' \emph{IEEE Trans. Inf. Theory}, vol.~67, no.~5, pp. 2739--2757,
  2021.

\bibitem{9716111}
------, ``On the decoding performance of spatially coupled {LDPC} codes with
  sub-block access,'' \emph{IEEE Trans. Inf. Theory}, vol.~68, no.~6, pp.
  3700--3718, 2022.

\bibitem{6325197}
A.~{Yedla}, Y.~{Jian}, P.~S. {Nguyen}, and H.~D. {Pfister}, ``A simple proof of
  threshold saturation for coupled scalar recursions,'' in \emph{Proc. Int.
  Symp. Turbo Codes Iterative Inf. Process (ISTC)}, 2012, pp. 51--55.

\bibitem{6887298}
A.~Yedla, Y.-Y. Jian, P.~S. Nguyen, and H.~D. Pfister, ``A simple proof of
  maxwell saturation for coupled scalar recursions,'' \emph{IEEE Trans. Inf.
  Theory}, vol.~60, no.~11, pp. 6943--6965, 2014.

\bibitem{7426827}
A.~Giurgiu, N.~Macris, and R.~Urbanke, ``Spatial coupling as a proof technique
  and three applications,'' \emph{IEEE Trans. Inf. Theory}, vol.~62, no.~10,
  pp. 5281--5295, 2016.

\bibitem{8002601}
S.~Moloudi, M.~Lentmaier, and A.~{Graell i Amat}, ``Spatially coupled
  turbo-like codes,'' \emph{IEEE Trans. Inf. Theory}, vol.~63, no.~10, pp.
  6199--6215, Oct. 2017.

\bibitem{7954697}
Y.~Y. Jian, H.~D. Pfister, and K.~R. Narayanan, ``Approaching capacity at high
  rates with iterative hard-decision decoding,'' \emph{IEEE Trans. Inf.
  Theory}, vol.~63, no.~9, pp. 5752--5773, Sep. 2017.

\bibitem{7430313}
I.~Andriyanova and A.~Graell~i Amat, ``Threshold saturation for nonbinary
  {SC-LDPC} codes on the binary erasure channel,'' \emph{IEEE Trans. Inf.
  Theory}, vol.~62, no.~5, pp. 2622--2638, 2016.

\bibitem{hsieh_2021}
\BIBentryALTinterwordspacing
K.~Hsieh, ``Spatially coupled sparse regression codes for single- and
  multi-user communications,'' Ph.D. dissertation, Cambridge University,
  Cambridge, UK, 2021. [Online]. Available:
  \url{https://www.repository.cam.ac.uk/handle/1810/323267}
\BIBentrySTDinterwordspacing

\bibitem{7541672}
C.~Häger, H.~D. Pfister, A.~Graell~i Amat, and F.~Brännström,
  ``Deterministic and ensemble-based spatially-coupled product codes,'' in
  \emph{Proc. IEEE Int. Symp. Inf. Theory (ISIT)}, 2016, pp. 2114--2118.

\bibitem{6120387}
A.~Yedla, P.~S. Nguyen, H.~D. Pfister, and K.~R. Narayanan, ``Universal codes
  for the {Gaussian MAC} via spatial coupling,'' in \emph{Proc. Allerton
  Conf.}, 2011, pp. 1801--1808.

\bibitem{9611454}
M.~M. Mashauri, A.~Graell~i Amat, and M.~Lentmaier, ``On the universality of
  spatially coupled {LDPC} codes over intersymbol interference channels,'' in
  \emph{IEEE Inf. Theory Workshop (ITW)}, 2021, pp. 1--6.

\bibitem{Moloudi_sctc_distance}
S.~Moloudi, M.~Lentmaier, and A.~{Graell i Amat}, ``Spatially coupled
  turbo-like codes: A new trade-off between waterfall and error floor,''
  \emph{IEEE Trans. Commun.}, vol.~67, no.~5, pp. 3114--3123, 2019.

\bibitem{6262475}
D.~G. Mitchell, A.~E. Pusane, and D.~J. Costello, ``Minimum distance and
  trapping set analysis of protograph-based {LDPC} convolutional codes,''
  \emph{IEEE Trans. Inf. Theory}, vol.~59, no.~1, pp. 254--281, 2013.

\bibitem{Battaglioni2024sc}
M.~Battaglioni, M.~Baldi, and F.~Chiaraluce, ``Bounds on the free distance of
  periodically time-varying {SC-LDPC} codes,'' \emph{IEEE Trans. Inf. Theory},
  vol.~70, no.~4, pp. 2419--2429, 2024.

\bibitem{7086074}
P.~M. Olmos and R.~L. Urbanke, ``A scaling law to predict the finite-length
  performance of spatially-coupled {LDPC} codes,'' \emph{IEEE Trans. Inf.
  Theory}, vol.~61, no.~6, pp. 3164--3184, 2015.

\bibitem{10068559}
R.~Sokolovskii, A.~{Graell i Amat}, and F.~Brännström, ``Finite-length
  scaling of {SC-LDPC} codes with a limited number of decoding iterations,''
  \emph{IEEE Trans. Inf. Theory}, vol.~69, no.~8, pp. 4869--4888, 2023.

\bibitem{9165839}
M.~Zhu, D.~G.~M. Mitchell, M.~Lentmaier, D.~J. Costello, and B.~Bai, ``Error
  propagation mitigation in sliding window decoding of braided convolutional
  codes,'' \emph{IEEE Transactions on Communications}, vol.~68, no.~11, pp.
  6683--6698, 2020.

\bibitem{10243110}
M.~Zhu, D.~G.~M. Mitchell, M.~Lentmaier, and D.~J. Costello, ``Error
  propagation mitigation in sliding window decoding of spatially coupled {LDPC}
  codes,'' \emph{IEEE Journal on Selected Areas in Information Theory}, vol.~4,
  pp. 470--486, 2023.

\bibitem{9112247}
S.~Mo, L.~Chen, D.~J. Costello, D.~G.~M. Mitchell, R.~Smarandache, and J.~Qiu,
  ``Designing protograph-based quasi-cyclic spatially coupled {LDPC} codes with
  large girth,'' \emph{IEEE Trans. Commun.}, vol.~68, no.~9, pp. 5326--5337,
  Sept. 2020.

\bibitem{9354189}
M.~Battaglioni, F.~Chiaraluce, M.~Baldi, and M.~Lentmaier, ``Girth analysis and
  design of periodically time-varying {SC-LDPC} codes,'' \emph{IEEE Trans. Inf.
  Theory}, vol.~67, no.~4, pp. 2217--2235, Apr. 2021.

\bibitem{1053951}
J.~Robinson and A.~Bernstein, ``A class of binary recurrent codes with limited
  error propagation,'' \emph{IEEE Trans. Inf. Theory}, vol.~13, no.~1, pp.
  106--113, Jan. 1967.

\bibitem{Alfarano2020ConstructionOL}
G.~N. Alfarano, J.~Lieb, and J.~Rosenthal, ``Construction of {LDPC}
  convolutional codes via difference triangle sets,'' \emph{Designs, Codes, and
  Cryptogr.}, vol.~89, pp. 2235--2254, Oct. 2020.

\bibitem{6510023}
A.~Gruner and M.~Huber, ``Low-density parity-check codes from transversal
  designs with improved stopping set distributions,'' \emph{IEEE Trans.
  Commun.}, vol.~61, no.~6, pp. 2190--2200, Jun. 2013.

\bibitem{10197554}
F.~Wang, S.~Zhao, J.~Wen, S.~Wang, and Z.~Li, ``A genie-aided approach to error
  floor estimation for spatially coupled serially concatenated codes,''
  \emph{IEEE Trans. Commun.}, vol.~71, no.~10, pp. 5713--5725, 2023.

\bibitem{sctc_convPerm}
M.~U. Farooq, A.~{Graell i Amat}, and M.~Lentmaier, ``Spatially-coupled
  serially concatenated codes with periodic convolutional permutors,'' in
  \emph{International Symposium on Topics in Coding (ISTC)}, 2021, pp. 1--5.

\bibitem{9732353}
H.-Y. Kwak, J.-W. Kim, H.~Park, and J.-S. No, ``Optimization of {SC-LDPC} codes
  for window decoding with target window sizes,'' \emph{IEEE Trans. Commun.},
  vol.~70, no.~5, pp. 2924--2938, 2022.

\bibitem{10802915}
O.~Griebel, B.~Hammoud, and N.~Wehn, ``Adaptive sliding window decoding of
  spatially coupled low-density parity-check codes: Algorithms and energy
  efficient implementations,'' \emph{IEEE Access}, vol.~12, pp.
  191\,140--191\,161, 2024.

\bibitem{8894067}
H.-Y. Kwak, D.-Y. Yun, and J.-S. No, ``Rate-loss mitigation of {SC-LDPC} codes
  without performance degradation,'' \emph{IEEE Trans. Commun.}, vol.~68,
  no.~1, pp. 55--65, 2020.

\bibitem{8419715}
R.~El-Khatib and N.~Macris, ``The velocity of the propagating wave for
  spatially coupled systems with applications to {LDPC} codes,'' \emph{IEEE
  Trans. Inf. Theory}, vol.~64, no.~11, pp. 7113--7131, 2018.

\bibitem{Yang2025spatiallycoupled}
\BIBentryALTinterwordspacing
S.~Yang and R.~Calderbank, ``Spatially-{C}oupled {QLDPC} {C}odes,''
  \emph{{Quantum}}, vol.~9, p. 1693, Apr. 2025. [Online]. Available:
  \url{https://doi.org/10.22331/q-2025-04-07-1693}
\BIBentrySTDinterwordspacing

\bibitem{6171810}
D.~Truhachev, ``Achieving {AWGN} multiple access channel capacity with spatial
  graph coupling,'' \emph{IEEE Commun. Lett.}, vol.~16, no.~5, pp. 585--588,
  2012.

\bibitem{8638808}
D.~Truhachev and C.~Schlegel, ``Coupling data transmission for multiple-access
  communications,'' \emph{IEEE Trans. Inf. Theory}, vol.~65, no.~7, pp.
  4550--4574, 2019.

\end{thebibliography}

\section{Biography}
\begin{IEEEbiographynophoto}{Min Qiu} (Senior Member, IEEE) received his Ph.D. degree in Electrical Engineering from the University of New South Wales (UNSW), Sydney, Australia, in 2019. From 2019 to 2025, he was a Postdoctoral Research Associate with UNSW. Since September 2025, he has been with Global College, Shanghai Jiao Tong University, Shanghai, China, where he is currently an Assistant Professor, and since April 2026, with its Terahertz Wireless Communications Interdisciplinary Research Center. He is the lead inventor of a patent on spatially coupled product-like codes. He is currently serving as an Associate Editor for IEEE Transactions on Communications and the Lead Guest Editor for Entropy Special Issue on Next-Generation Channel Coding. His main research interests are in the field of channel coding and communication theory. 
\end{IEEEbiographynophoto}

\begin{IEEEbiographynophoto}{Xiaowei Wu} (Member, IEEE) received her B.E., M.Sc., and Ph.D. degrees from the University of New South Wales, Sydney, Australia, in 2014, 2017, and 2022, respectively. From 2022 to 2025, she was a postdoctoral research assistant with the Technology and Engineering Center for Space Utilization (CSU), Chinese Academy of Sciences. She is currently an associate research fellow at CSU. Her research interests include channel coding and free space optical communications.
\end{IEEEbiographynophoto}

\begin{IEEEbiographynophoto}{Peng Kang} (Senior Member, IEEE) received the B.E. degree in communications engineering from Fuzhou University, Fuzhou, China, in 2013, the M.Sc. degree in telecommunications from the Hong Kong University of Science and Technology (HKUST), Hong Kong SAR, in 2014, and the Ph.D. degree in electrical engineering from the University of New South Wales (UNSW), Australia, in 2019. From 2020 to 2023, he was a Post-Doctoral Research Fellow with Singapore University of Technology and Design (SUTD), Singapore. He is currently an Associate Professor with Fuzhou University, Fuzhou, China. His research interests include channel coding, iterative decoding techniques, and their application in communication systems.
\end{IEEEbiographynophoto}

\begin{IEEEbiographynophoto}{Lei Yang} (Member, IEEE) is a professor at the Technology and Engineering Center for Space Utilization (CSU), Chinese Academy of Sciences. He received the Ph.D. degree from the Beijing Institute of Technology in 2015. His research interests include single-photon deep-space optical communications, error control coding, and iterative signal processing.
\end{IEEEbiographynophoto}

\begin{IEEEbiographynophoto}{Jinhong Yuan} (Fellow, IEEE) received the B.E. and Ph.D. degrees in electronics engineering in 1991 and 1997, respectively. From 1997 to 1999, he was a Research Fellow with the School of Electrical Engineering, University of Sydney, Sydney, Australia. In 2000, he joined the School of Electrical Engineering and Telecommunications, University of New South Wales, Sydney, Australia, where he is currently Head of School. He has published two books, five book chapters, over 300 papers in telecommunications journals and conference proceedings, and 50 industrial reports. He is a co-inventor of one patent on MIMO systems and four patents on low-density parity-check codes. He has co-authored five Best Paper Awards and one Best Poster Award, including 2025 IEEE Asia-Pacific Best Paper Award, the Best Paper Award from the IEEE International Conference on Communications, Kansas City, USA, in 2018, the Best Paper Award from IEEE Wireless Communications and Networking Conference, Cancun, Mexico, in 2011, and the Best Paper Award from the IEEE International Symposium on Wireless Communications Systems, Trondheim, Norway, in 2007. He is an IEEE Fellow and listed as a 2025 Highly-Cited Researcher. He served as the IEEE NSW Chapter Chair of Joint Communications/Signal Processions/Ocean Engineering Chapter during 2011-2014 and served as an Associate Editor for the IEEE Transactions on Communications during 2012-2017 and 2020-2025 and IEEE Transactions on Wireless Communications during 2019-2024. His current research interests include error control coding and information theory, communication theory, wireless communications, and delay-Doppler domain signal processing and communications.
\end{IEEEbiographynophoto}

\vfill

\end{document}